\documentclass[a4paper,aps,pra,reprint,superscriptaddress,onecolumn,notitlepage,11pt]{revtex4-1} 

\usepackage[utf8]{inputenc} 
\usepackage{textcomp} 
\usepackage{graphicx}  
\usepackage{mathrsfs,amsmath,amssymb}  
\usepackage{bm}  
\usepackage{rotating}
\usepackage{color}

\begin{document}
   
\title{Modulation Instability Induced Frequency Comb Generation\\
in a Continuously Pumped Optical Parametric Oscillator}

\author{S.~Mosca}
\email{simona.mosca@ino.it}
\affiliation{CNR-INO, Istituto Nazionale di Ottica, Via Campi Flegrei 34, I-80078 Pozzuoli (NA), Italy}

\author{M.~Parisi}
\affiliation{CNR-INO, Istituto Nazionale di Ottica, Via Campi Flegrei 34, I-80078 Pozzuoli (NA), Italy}
\author{I.~Ricciardi}

\affiliation{CNR-INO, Istituto Nazionale di Ottica, Via Campi Flegrei 34, I-80078 Pozzuoli (NA), Italy}
\affiliation{INFN, Istituto Nazionale di Fisica Nucleare, Sez. di Napoli, Complesso Universitario di M.S. Angelo, Via Cintia, Napoli, 80126 Italy}

\author{F. Leo}
\affiliation{OPERA-photonics, Universit\'e libre de Bruxelles, 50 Avenue F. D. Roosevelt, CP 194/5, B-1050 Bruxelles, Belgium}

\author{T. Hansson}
\affiliation{Dipartimento di Ingegneria dell'Informazione, Universit\`a di Brescia, Via Branze 38, I-25123 Brescia, Italy}

\author{M. Erkintalo}
\affiliation{The Dodd-Walls Centre for Photonic and Quantum Technologies, Department of Physics, The University of Auckland, Auckland 1142, New Zealand}

\author{P.~Maddaloni}
\affiliation{CNR-INO, Istituto Nazionale di Ottica, Via Campi Flegrei 34, I-80078 Pozzuoli (NA), Italy}
\affiliation{INFN, Istituto Nazionale di Fisica Nucleare, Sez. di Napoli, Complesso Universitario di M.S. Angelo, Via Cintia, Napoli, 80126 Italy}

\author{P.~De~Natale}
\affiliation{CNR-INO, Istituto Nazionale di Ottica, Largo E. Fermi 6, I-50125 Firenze, Italy}

\author{S. Wabnitz}
\affiliation{Dipartimento di Ingegneria dell'Informazione, Universit\`a di Brescia, and CNR--INO, Via Branze 38, I-25123 Brescia, Italy}
\affiliation{Novosibirsk State University, 1 Pirogova str., Novosibirsk 630090, Russia}

\author{M.~De~Rosa}
\email{maurizio.derosa@ino.it}
\affiliation{CNR-INO, Istituto Nazionale di Ottica, Via Campi Flegrei 34, I-80078 Pozzuoli (NA), Italy}
\affiliation{INFN, Istituto Nazionale di Fisica Nucleare, Sez. di Napoli, Complesso Universitario di M.S. Angelo, Via Cintia, Napoli, 80126 Italy}


\begin{abstract}
Continuously pumped passive nonlinear cavities can be harnessed for the creation of novel optical frequency combs. While most research has focused on third-order ``Kerr'' nonlinear interactions, recent studies have shown that frequency comb formation can also occur via second-order nonlinear effects. 
Here, we report on the formation of quadratic combs in optical parametric oscillator (OPO) configurations. Specifically, we demonstrate that optical frequency combs can be generated in the parametric region around half of the pump frequency in a continuously-driven OPO. 
We also model the OPO dynamics through a single time-domain mean-field equation, identifying previously unknown dynamical regimes, induced by modulation instabilities, which lead to comb formation.
Numerical simulation results are in good agreement with experimentally observed spectra. 
Moreover, the analysis of the coherence properties of the simulated spectra shows the existence of correlated and phase-locked combs.
Our results reveal previously unnoticed dynamics of an apparently well assessed optical system, and can lead to a new class of frequency comb sources that may stimulate novel applications by enabling straightforward access to elusive spectral regions, such as the mid-infrared.
\end{abstract}

\maketitle

\newpage


Optical frequency comb (OFC) generation in passive nonlinear cavities has attracted increasing interest over the last decade as an alternative way to the traditional techniques based on femtosecond mode-locked lasers~\cite{Holzwarth:2000aa,Jones:2000tn}. 
The vast majority of studies have focused on comb formation via third-order $\chi^{(3)}$ nonlinear interactions~\cite{DelHaye:2007gi,Chembo:2010ii,Hansson:2013jy,Kippenberg:2011fc}.
More recently, however, OFCs have also been demonstrated in continuously driven cavities with quadratic $\chi^{(2)}$ nonlinearities~\cite{Ulvila:2013jv,Ulvila:2014bx,Ricciardi:2015bw,Mosca:2015wh}. 
In particular, OFCs can be generated in a cavity with a $\chi^{(2)}$ nonlinear crystal that is phase-matched for second harmonic generation (SHG)~\cite{Ricciardi:2015bw,Mosca:2015wh}. When the SHG cavity is pumped beyond the threshold for the onset of a so-called internally pumped optical parametric oscillator (IP-OPO)~\cite{Schiller:1996gx,Schiller:1997dp,White:1997ta}, other cascaded $\chi^{(2)}$ processes occur, eventually leading to a comb emission around both the fundamental and second harmonic frequencies.

A particularly attractive feature of quadratic resonators is that they can permit the generation of frequency combs in spectral regions far from the pump frequency. For example, phase-matched SHG can allow for the generation of visible frequency combs using near-infrared pump lasers. On the other hand, numerical simulations~\cite{Hansson:2016kz} suggest that the inverse process of parametric down-conversion could similarly allow for direct generation of frequency combs in the elusive mid-IR region. 
Conversely, as Kerr microresonators can only generate combs around the pump wavelength, the scarcity of mid-IR continuous-wave (cw) lasers and appropriate nonlinear crystals prevents that technology from being directly applied in the mid-IR region. 
Applications such as molecular spectroscopy~\cite{Schliesser:2012dn,Maddaloni:2009bg} would greatly benefit  from a technology that allows for more straightforward and efficient generation of mid-IR frequency combs than existing solutions, based on the nonlinear transfer of near-IR OFCs in a separate stage~\cite{Maddaloni:2006ka,Erny:2007dj,Sun:2007wu,Adler:2009ka,Wong:2010gw,Leindecker:2011ch,Leindecker:2012ed,Keilmann:2012bb,Galli:2013cg,Gambetta:2013ci,Ru:2017db}.

Although the prospect of generating frequency combs through direct intracavity parametric down-conversion is attractive, no experimental demonstration has been reported, so far. 
The possibility of generating several new frequency components in a singly resonant, nondegenerate OPO was theoretically predicted already in 1969~\cite{Kreuzer:1969vm}.
Well before the first clear demonstration of degenerate OPO emission~\cite{Nabors:1990vl}, several theoretical studies have  been devoted to the description of cw pumped OPOs where both the pump and the degenerate parametric fields resonate~\cite{Drummond:1980bj,Lugiato:1988ep}, revealing a rich variety of dynamical regimes. However, these pioneering modelling efforts neglected both diffraction and dispersion.
Models that include the effect of diffraction predict the formation of transverse spatial structures, such as roll patterns or solitons~\cite{Oppo:1994en,Oppo:1994kf,DeValcarcel:1996df,Staliunas:1995ky,Longhi:1996ft,Longhi:1997iy}.
The existence of stable pulsed states (solitons or temporal patterns) has meanwhile been predicted by including the effects of group velocity dispersion or spectral filtering in the model equations~\cite{Longhi:1995jm,Longhi:1996dh,Longhi:1996dm,Longhi:1996gt}.
Of course, the spectral counterpart of such temporal structures would constitute a frequency comb.

Here, we experimentally demonstrate generation of quadratic frequency combs in a cw pumped, degenerate, doubly resonant OPO. 
Significantly, we observe comb emission both around the pump angular frequency $2\omega_0$ and in the parametric spectral range around the degeneracy frequency $\omega_0$. 
We also present a time domain theoretical model for the intracavity field dynamics that includes the effects of cavity dispersion, and we derive a single mean-field equation that governs the evolution of the slowly-varying field amplitude at the parametric frequency.
We find that the inclusion of dispersion induces a modulation instability of both the constant solutions predicted by the usual, dispersionless OPO model~\cite{Boyd:NLO}, and can be responsible for the formation of comb spectral structures.

To model comb dynamics in our OPO system, we consider an optical parametric oscillator pumped by the field $B_\mathrm{in}$ at $2\omega_0$, which operates near degeneracy, so that signal and idler fields can be described by a single field envelope $A$. 
We assume that the cavity of length $L$ is filled with the nonlinear medium, and is resonant only for parametric waves, while the cw pump beam $B_\mathrm{in}$ leaves the cavity after each round-trip.  
We also assume that the wave vectors $k_1$ and $k_2$ at $\omega_0$ and $2\omega_0$, respectively, satisfy the degenerate phase matching condition $\Delta k = 2 k_1 - k_2=0$.

OPO dynamics can be described by an infinite dimensional map for the field amplitudes. 
The map consists of coupled propagation equations that describe the evolution of the field amplitudes at $\omega_0$ and $2\omega_0$ over one cavity round-trip, as well as boundary equations that describe the input/output relations for the fields at the beginning of each round-trip.
Following the approach of Ref.~\cite{Leo:2016kj}, the infinite dimensional map can be combined into a single mean-field equation for the parametric field $A$  \footnote{Supplemental Material at ... includes the detailed derivation and the stability analysis.}\setcounter{footnote}{16}
\begin{align}
  t_\mathrm{R}& \frac{\partial A(t,\tau)}{\partial t}  = 
   \left[ -\alpha_1 - i \delta_1 -i\frac{L{k}_1''}{2}\frac{\partial^2}{\partial \tau^2} \right] \, A 
   - \mu^2 A^*\left[A^2(t,\tau)\otimes I(\tau)\right] 
   + i \mu  B_\mathrm{in} A^* \, e^{-i \xi} \, \mathrm{sinc} (\xi) .
 \label{MFE}
\end{align}
Here, 
$t$ is a ``slow time'' variable that describes the envelope's evolution over successive round-trips, while the “fast-time” $\tau$ describes the temporal profiles of the fields in a reference frame moving with the group velocity of light at $\omega_0$,   
$\alpha_1$ are the total cavity linear losses, 
$\delta_{1} \simeq  (\omega_0- \omega_c) t_\mathrm{R}$ is the phase detuning between the parametric field and the closest cavity resonance with frequency $\omega_\mathrm{c}$, $t_\mathrm{R}$ is the round-trip time, 
${k}''_{1} = \mathrm{d}^2k/\mathrm{d}\omega^2|_{\omega_0}$ is the group velocity dispersion coefficient at $\omega_0$, 
$\mu=\kappa L$, where $\kappa$ is the nonlinear coupling constant, 
and $\xi=\Delta k L/ 2$ is the wave vector mismatch parameter.
The symbol $\otimes$ denotes the convolution operator and  $I(\tau)= {\cal F}^{-1}[\hat{I}(\Omega)]$ is a nonlinear response function, where $\hat{I}(\Omega) = (1-i x - e^{-ix})/x^2$,
with $x \equiv x(\Omega) = [\Delta k + i\hat{k}(\Omega)]L$ and $\hat{k}(\Omega)  = -\alpha_{c,2}/2+i\left[\Delta{k}'\Omega+({k}_2''/2)\Omega^2\right]$. In the last expressions, the spectral frequency $\Omega$ is the offset angular frequency with respect to $\omega_0$, 
$\Delta {k}' = \mathrm{d}k/\mathrm{d}\omega|_{2\omega_0}-\mathrm{d}k/\mathrm{d}\omega|_{\omega_0}$ is the group-velocity mismatch, or temporal walk-off, between the fields at $\omega_0$ and $2\omega_0$, 
$\alpha_{c2}$ are the propagation losses at $2\omega_0$, 
and ${k}''_{2} = \mathrm{d}^2k/\mathrm{d}\omega^2|_{2\omega_0}$ is the group velocity dispersion coefficient at $2\omega_0$.

\begin{figure}[t]
\begin{center}
\includegraphics*[viewport=0 0 610 300, clip, width=\columnwidth]{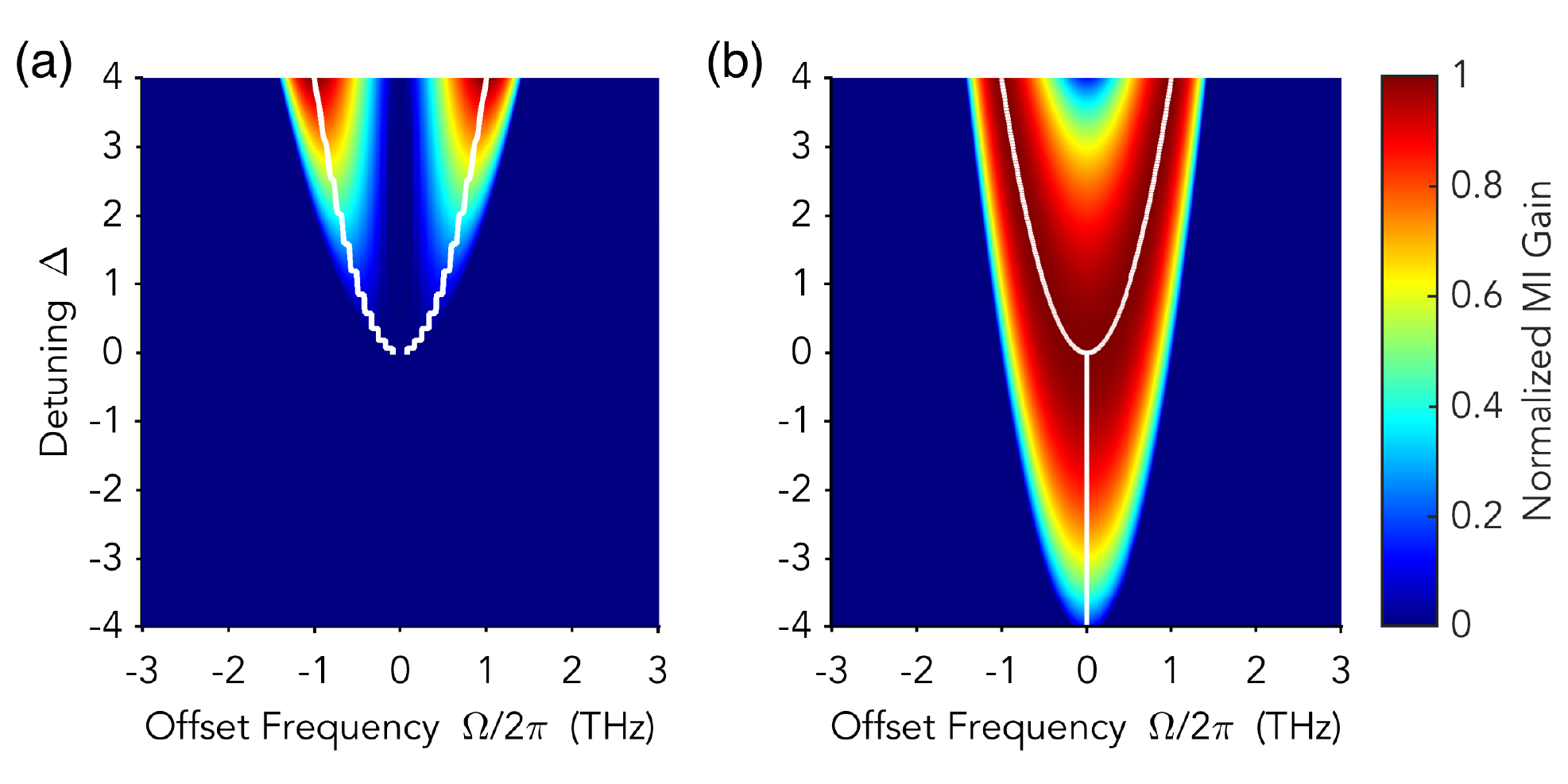}
\caption{ 
MI gain of (a) the non-zero constant solution and (b) trivial zero solution, as a function of the offset frequency $\Omega/2\pi$ and the detuning $\Delta$, calculated for our experimental parameters: 
$\alpha_1=0.017$, 
$L=15$~mm, 
$k_1^{\prime\prime}=0.234$~ps$^2$/m, 
$k_2^{\prime\prime}=0.714$~ps$^2$/m,  
$\Delta  k^\prime=792$~ps/m, 
$\kappa=6.58$~$\mathrm{W}^{-1/2}\mathrm{m}^{-1}$, 
$\xi=0$,
$P_\mathrm{in}=300$~mW. White curves highlight the maxima of the instability gain. 
}
\label{Fig1}
\end{center}
\end{figure}

Equation~(\ref{MFE}) is a generalization of the dissipative nonlinear Schr\"odinger equation, with a parametric driving term~\cite{Longhi:1996dm,Longhi:1996dh} and  a non-instantaneous interaction term~\cite{Kibler:2012ho,Hansson:2014do,Okawachi:2017jl}.
The function $I(\tau)$ is the same  nonlinear response kernel which describes singly resonant cavity SHG~\cite{Leo:2016kj,Hansson:2017cs}: physically, it is related to the nonlinear losses and phase shifts due to cascaded sum frequency generation process.
Equation~(\ref{MFE}) has a trivial zero solution, $A=0$, and a nontrivial time independent solution, $A_0=|A_0| \, e^{i\phi}$.
Both the trivial and non-zero solution can exhibit modulation instability (MI) gain, which reinforces random fluctuations and leads to the exponential growth of sidebands around the carrier frequency $\omega_0$, and eventually to the formation of frequency combs (Supplemental Material, Sec.~2 \cite{Note1}). 
Figure~\ref{Fig1} shows the MI gain for the constant finite and zero solutions, respectively, as a  function of the offset frequency $\Omega/2\pi$ and the relative detuning $\Delta=\delta_1/\alpha_1$. 
For positive detunings, both solutions display MI gain, with frequency symmetric pairs of gain maxima. 
For $\Delta>0$, the critical frequency $\Omega_\mathrm{max}$ for which the MI gain is maximum  (white curves in Fig.~\ref{Fig1}) nicely follows the relation $\Omega_\mathrm{max}^2 =  (2/L k''_1)\delta_1$, analogously to the spatial case of transverse pattern formation in  different nonlinear optical systems~\cite{Staliunas:2000tu,Bortolozzo:2001dt,EstebanMartin:2004dv}.
For zero and negative detunings, the zero solution still exhibits MI gain, with peak gain at  $\Omega=0$.
It is worth mentioning that, as in SHG systems, the walk-off $\Delta k'$ plays a key role in  determining the MI gain profile associated with the non-zero solution. 
Indeed, for the parameters quoted in Fig.~\ref{Fig1}, MI only manifests itself for relatively high walk-off values, while it is absent for zero walk-off (Supplemental Material, Fig.~S1 \cite{Note1}). 
In contrast, the instability of the zero solution, which is not expected in the usual dispersionless analysis of the doubly resonant OPO, does not depend on the walk-off, but is rather induced by group-velocity dispersion.
A similar dynamic instability arises in the spatial case, when diffraction is considered~\cite{Oppo:1994en}.

\begin{figure}[pt]
\begin{center}
\includegraphics*[viewport=0 20 880 690, clip, width=\columnwidth]{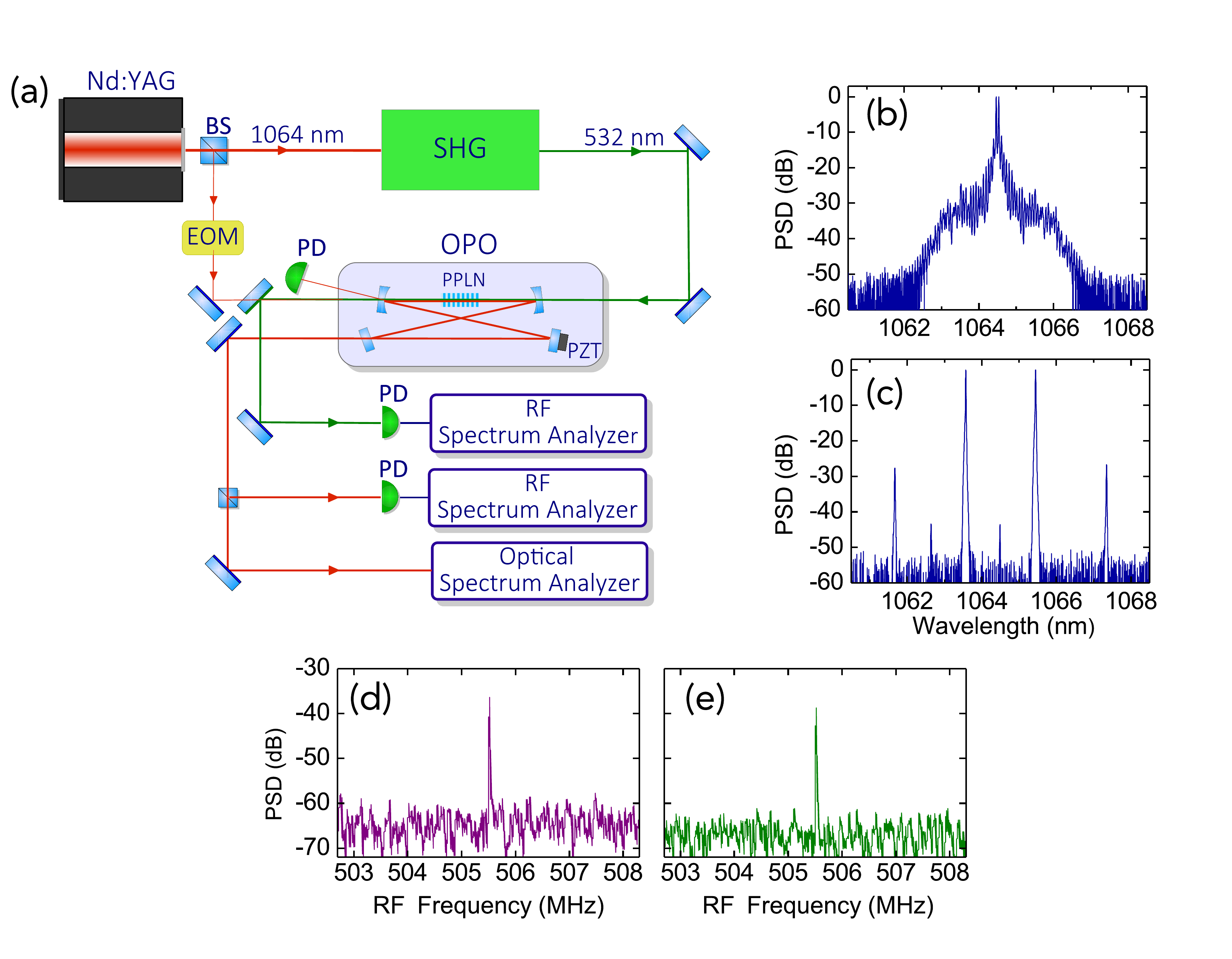}
\caption{
(a) Schematic of the experimental setup.
BS: beam splitter; EOM: electro-optic phase modulator; SHG: second harmonic generation cavity; OPO: optical parametric cavity; PPLN: periodically poled lithium niobate crystal; PZT: piezoelectric actuator; PD: photodiode. 
(b) and (c), experimental infrared comb spectra with zero cavity detuning for different values of the pump power (resolution bandwidth, $\mathrm{RBW}= 6$~GHz). (d) and (e), RF spectra showing beat notes at 505~MHz,  in the IR and visible region, respectively, corresponding to the comb spectrum shown in (b) ($\mathrm{RBW}=1$~kHz). 
}
\label{Fig2}
\end{center}
\end{figure}


 Our experiment is based on a degenerate OPO pumped by a frequency doubled cw Nd:YAG laser [see Fig. \ref{Fig2}(a)]. 
  The OPO is based on a temperature-stabilized, 15-mm-long, periodically-poled 5\%-MgO-doped lithium niobate crystal (PPLN), with a grating period $\Lambda = 6.92$~$\mu$m,  enclosed in a bow-tie cavity resonating for the parametric wavelengths around 1064~nm (further details are given in Supplemental Material, Sec.~3 \cite{Note1}).
The cavity has a free spectral range (FSR) of $505 \pm 1$ MHz, and a finesse of 180, the linear propagation losses at $\omega_0$ being less than 1\% over one round-trip.
To make sure that the OPO cavity is resonant at the degeneracy frequency $\omega_0$, the cavity is frequency locked by the Pound--Drever--Hall (PDH) technique~\cite{Drever:1983gx}, using the phase modulated IR laser light reflected by a beam splitter, which is injected in the OPO cavity in a direction opposite to the SH pump beam. 
The PDH error signal is processed by an electronic servo control and fed to the cavity piezoelectric actuator that moves one of the cavity mirrors.
 The laser, emitting 2~W at 1064~nm (Innolight, Mephisto 2000), is frequency doubled in a periodically poled lithium tantalate (PPLT) crystal placed in an external SHG cavity~\cite{Ricciardi:2010kd}.
We determined the phase-matching temperature for degenerate operation of the OPO to be $T = (61.0 \pm 0.1)^\circ$C by maximizing the efficiency of the SHG when pumped directly with the Nd:YAG laser. 

 The spectral composition of the parametric waves is analyzed by an optical spectrum analyzer (OSA), while two fast photodiodes monitor the light coupled out of the cavity in the respective optical ranges around $\omega_0$ and $2\omega_0$. 
 We note that, because of its limited spectral range (600--1700~nm), our OSA is not capable of detecting the comb around the SH frequency.
Rather, evidence of SH comb is inferred from the radio-frequency (RF) spectrum of the photodiode signal.


The typical power threshold for the onset of degenerate single-frequency OPO emission is around 30~mW.
By further increasing the pump power to about 85~mW, we observe the generation of a stable optical frequency comb.
Figure~\ref{Fig2}(b) shows the spectra detected by pumping the OPO with $P_\mathrm{in}=300$~mW of green light, corresponding to an integrated IR power of 120~mW for the signal field at the cavity output. 
In this case, we detected narrow 1-kHz-resolution-limited beat notes at 1 FSR in the IR and visible, as shown in Fig.~\ref{Fig2}(d) and (e), respectively.
This indicates a comb structure with lines equally spaced by one cavity FSR (at least, within the 1~kHz instrumental resolution) both around the pump frequency $2\omega_0$ and the parametric frequency $\omega_0$.
 When $P_\mathrm{in}=520$~mW, the IR comb emission changes to a few teeth around the degeneracy point, with a spacing of about 500~GHz, as shown in Fig.~\ref{Fig2}(c). In this case, the absence of beat notes within the bandwidth of the fast detectors suggests that the spectral peaks are indeed isolated single lines.  
As the pump power further increases, the stability of the OPO cavity locking deteriorates, and the error signal starts to oscillate erratically, thus precluding the observation of stable comb emission.

\begin{figure}[pt]
\begin{center}
\includegraphics*[bbllx=0bp,bblly=0bp,bburx=580bp,bbury=710bp,width=0.99\columnwidth]{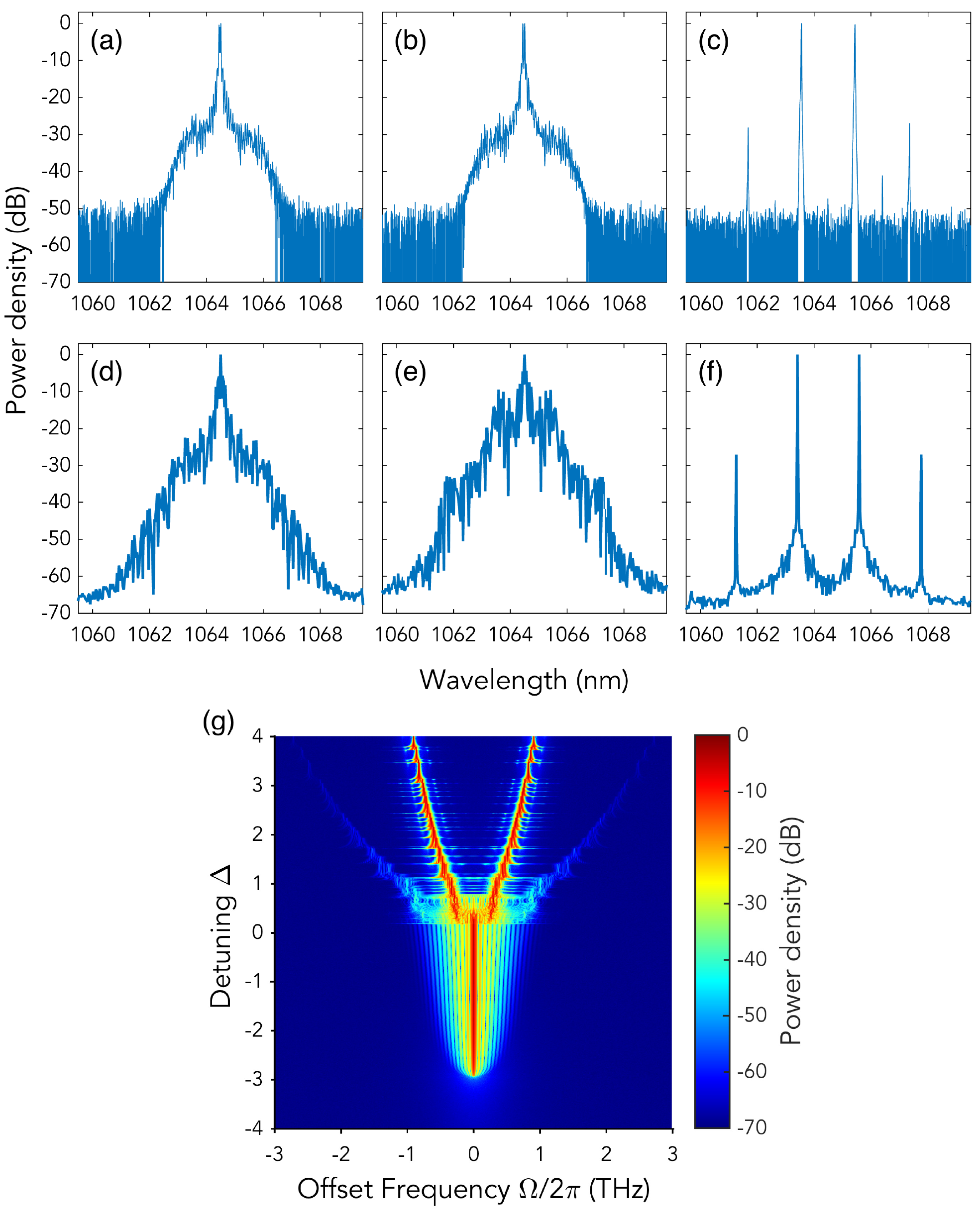}
\caption{
Normalized infrared OPO spectra for different cavity detunings. (a)-(c)   experimental spectra for relative detunings $\Delta = -0.3,\, 0,\, +0.3$, respectively. (d)-(f) corresponding numerically calculated spectra. 
(g) Map of calculated OPO spectra as a function of the offset from the degeneracy frequency and the relative detuning.  
Measurements and simulations refer to 300~mW of pump power. 
}
\label{Fig3}
\end{center}
\end{figure}

We also experimentally studied the effect of cavity detuning on the comb spectra.
To this end, we added a voltage offset to the error signal used for cavity locking, which allowed us to explore relative phase detunings  $\Delta$ between $-0.3$ and $0.3$.
Figure \ref{Fig3}(a)-(c) shows the experimental comb spectra recorded for  $\Delta =  -0.30, 0.00, +0.30$ ($\pm0.02$), respectively, for 300~mW of pump power. 
Corresponding spectra obtained from numerical simulations are shown in Fig.~\ref{Fig3}(d)-(f). 
Experimental spectra for negative and zero detunings are very similar, displaying a 1~FSR line spacing, while, for the positive detuning, the experimental spectrum consists of two pairs of symmetric lines, very similar to that observed in Fig.~\ref{Fig2}(c). We found a good agreement with the corresponding spectra, shown in Fig.~\ref{Fig3}(d)--(f), calculated by numerically integrating Eq.~(\ref{MFE}) over the slow time $t$ (Supplemental Material, Sec.~1 \cite{Note1}). 
All the simulation parameters are listed in the caption of Fig.~\ref{Fig1}. 
Numerical simulations show that the  spectrum of Fig.~\ref{Fig2}(c) is a still possible solution of the model of Eq.~(\ref{MFE}), but for a different set of parameters.
We believe, however, that the discrepancy can hardly be ascribed to imprecise knowledge of the parameters alone.
Rather, we speculate that additional effects, such as thermo-optical nonlinearities, might underlie the observed discordance, as well as the instability of the cavity locking at high pump power.

To give a more comprehensive picture of the comb dynamics, we simulated the evolution of the comb spectrum as function of $\Delta$ over a wider range of values than those accessible in our experiments. 
In this case, we swept the detuning from negative to positive values over $10^7$ round-trips.
As shown in Fig.~\ref{Fig3}(g), for negative and small positive detunings, we observe comb spectra with 1-FSR-spaced lines. 
Around $\Delta=0.3$ the spectra show a sharp transition to a different power distribution, but still with a 1 FSR mode spacing, which, as the detuning further increases ($\Delta \gtrsim 1$), evolves into small 1-FSR-spaced lines groups, which are widely-spaced from the degeneracy frequency. In general, the spectral dynamics is in good agreement with the expected MI gain. 
Finally, we notice that, while dispersion is essential for comb formation, it also entails a frequency dependent cavity FSR, which eventually sets a limit on the spectral bandwidth of the resulting combs, as we numerically checked (Supplemental Material, Fig.~S2 \cite{Note1}).

\begin{figure}[pt]
\begin{center}
\includegraphics*[viewport=0 0 570 340, clip,width=\columnwidth]{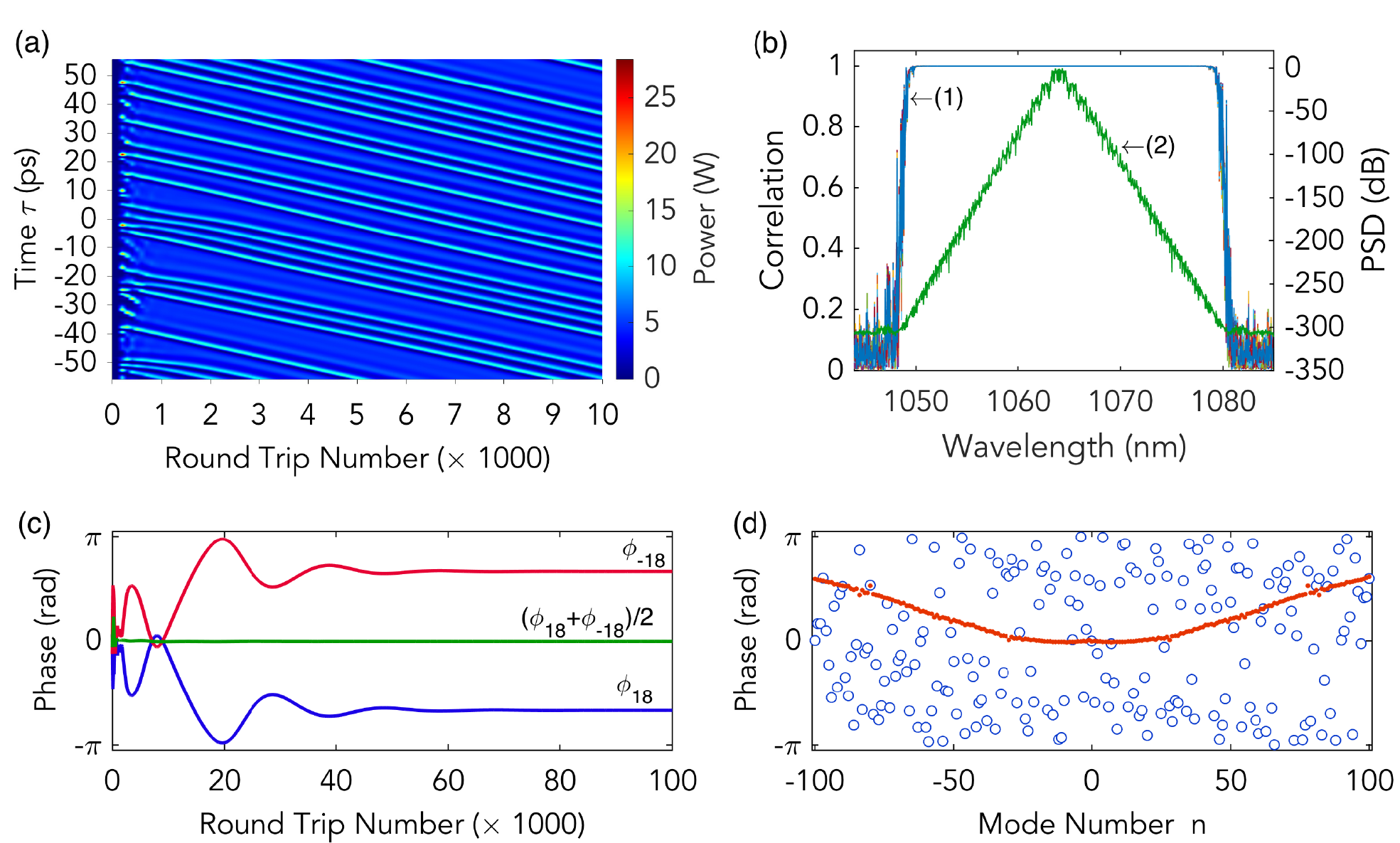}
\caption{ 
Coherence analysis.
(a) Temporal evolution of the parametric field envelope ($t_\mathrm{R}=111$~ps,  $P_\mathrm{in}=300$~mW and $\Delta=-2$). 
(b) Traces (1) are the first order correlation function $g^{(1)}(m,\lambda)$ as a function of wavelength, calculated for different time delays up to 100~$\mu$s; trace (2) is the comb spectrum profile. 
(c) Offset phase evolution for two symmetric modes $\pm18$ and their average phase.
(d) Steady state offset phase (blue open circles) and symmetric average phase (red dots), as a function of mode number $n$. 
 }
\label{Fig4}
\end{center}
\end{figure}

Numerical simulations also provide insights into the time-domain emission profile, which cannot be experimentally accessed due to the insufficient bandwidth of our photodetectors. 
Simulations of quadratic combs in cavity SHG have shown stable steady state solutions with repeatable and regular temporal patterns, such as Turing patterns or solitary waves, which exhibit phase locking
~\cite{Leo:2016kj,Leo:2016df,Hansson:2017cs,Parisi:2017fp}.
Interestingly, simulations of our OPO show that, in general, the intracavity temporal profiles do not correspond to regular patterns, and that their precise form depends sensitively on the random seed used as initial condition. 
Nevertheless, even though the temporal profiles do not correspond to clean pulse trains, the frequency combs can still exhibit a high degree of coherence for a wide range of pump powers and detunings.
Figure~\ref{Fig4}(a) shows the time evolution of the parametric field envelope over the first $10\,000$ round-trips ($\sim 1$~$\mu$s).
The field profile rapidly evolves to a steady state, slowly drifting in the frame of reference that moves at the group velocity of the light at $\omega_0$. 
As in SHG combs~\cite{Leo:2016kj,Leo:2016df,Hansson:2017cs}, the relative drift of the temporal pattern, much slower than the walk-off $\Delta k^\prime$, indicates a nonlinear compensation of the walk-off induced linear spectral phase, which can lead, in principle, to group-velocity-locked fundamental and parametric waves.

To quantify the coherence of this state, we calculated the first-degree correlation function~\cite{Erkintalo:2014fq}
$
g^{(1)}(m,\lambda) = |\langle \tilde{E}^*(t_j,\lambda) \tilde{E}(t_{j+m},\lambda) \rangle |/\langle | \tilde{E}(t_j,\lambda) |^2 \rangle ,
$
where
$\tilde{E}(t_j,\lambda)$ 
is the Fourier transform of the complex intracavity field, calculated at multiples of the round-trip time $t_j = j t_{R}$, with $j$  an integer number. Angle brackets denote an ensemble average.
As shown in Fig.~\ref{Fig4}(b), after reaching steady state, we obtained an almost perfect correlation $g^{(1)}(m,\lambda)=1$, over a 100~$\mu$s time scale ($m \simeq 10^6$), for all the comb teeth that emerge from the noise floor.  
For each comb mode $n$, we calculated the phase $\varphi_n$, as a function of the round-trip time. Figure~\ref{Fig4}(c) shows the evolution of the relative phase $\phi_n=\varphi_n-\varphi_0$ for two symmetric modes, $n=\pm18$. We notice that, on the time scale of several tens of thousands of $t_\mathrm{R}$, each phase settles to a steady value, which randomly depends on the initial seeding noise [open circles, Fig.~\ref{Fig4}(d)]. Nevertheless, the average phase of pairs of symmetric modes, e.g. $(\phi_{18}+\phi_{-18})/2$, settles to a  steady value on a shorter time scale (a few times the cavity photon lifetime), and exhibits a regular and reproducible trend, when we repeat  the same simulation with a different seeding random noise, as shown by the seemingly smooth distribution of red dots in Fig.~\ref{Fig4}(d). 
The rapid anti-symmetrization of the phase profile around the degeneracy frequency is essentially related to the intrinsic symmetry of the process that initiates the comb (modulation instability, in our case), and can eventually lead to comb mode phase locking~\cite{Wen:2016cz}.
It is noteworthy that an indirect experimental evidence of this intermodal phase coherence is provided by the narrow resolution-limited beat notes shown in Fig.~\ref{Fig2}(d) and (e).

In conclusion, we have  experimentally demonstrated frequency comb generation in a cw pumped degenerate OPO, and theoretically derived a mean-field equation, which rather well describes the observed dynamic regimes. 
Our findings disclose new aspects of OPO dynamics, never observed till now, and open an entirely new scenario in the field of nonlinear optics.
In fact, our theoretical framework can be extended to other OPO configurations, such as non-degenerate and triply resonant OPOs.
In view of the increasing number of applications of OFCs in many fields of Physics, our work paves the route to a new class of devices, which could efficiently provide OFCs in spectral regions, like the MIR, where comb sources are highly desirable but still difficult to get.
Exploiting the described comb generation scheme, a larger comb bandwidth can be enabled by tailoring the dispersion in chip-scale waveguide resonators that exhibit large intrinsic or induced second-order nonlinearity, through efficient quasi-phase matching schemes~\cite{Morais:2017va,Timurdogan:2017jg}.

\vskip 10pt
\noindent

The authors acknowledge financial support by: Ministero dell’Istruzione, dell’Universit\`a e della Ricerca (Projects PRIN-2015KEZNYM ``NEMO--Nonlinear dynamics of optical frequency combs'' and FIRB n. RBFR13QUVI “Optomechanical tailoring of squeezed light”); Ministero degli Affari Esteri e della Cooperazione Internazionale (Project ``NOICE Joint Laboratory''); and Rutherford Discovery Fellowships and Marsden fund of the Royal Society of New Zealand. The research leading to these results has received funding from the European Union's Horizon 2020 research and innovation programme under the Marie Sklodowska-Curie grant agreement No GA-2015-713694. The work of S.W.  is supported by Ministry of Education and Science of the Russian Federation (Minobrnauka) (14.Y26.31.0017).

\newpage

\setcounter{equation}{0}
\setcounter{figure}{0}
\setcounter{table}{0}
\setcounter{page}{1}
\makeatletter
\renewcommand{\theequation}{\arabic{equation}}
\renewcommand{\thefigure}{\arabic{figure}}
\renewcommand{\thetable}{\arabic{table}}
\renewcommand{\bibnumfmt}[1]{[#1]}
\renewcommand{\citenumfont}[1]{#1}

\thispagestyle{empty}

\begin{center}
{\large \bf Supplemental Material for:\\
``Modulation Instability Induced Frequency Comb Generation\\
in a Continuously Pumped Optical Parametric Oscillator''}
\vskip12pt

S.~Mosca$^1$,
M.~Parisi$^1$,
I.~Ricciardi$^{1,2}$,
F.~Leo$^3$,
T. Hansson$^4$,
M.~Erkintalo$^5$,
P.~Maddaloni$^{1,2}$,
P.~De~Natale$^6$,
S.~Wabnitz$^{7,8}$,
M.~De~Rosa$^{1,2}$

\vskip8pt
{\it\small
$^1$CNR-INO, Istituto Nazionale di Ottica, Via Campi Flegrei 34, I-80078 Pozzuoli (NA), Italy
\\
$^2$
INFN, Istituto Nazionale di Fisica Nucleare, Sez. di Napoli, 
\\Complesso Universitario di M.S. Angelo, Via Cintia, Napoli, 80126 Italy
\\
$^3$OPERA-photonics, Universit\'e Libre de Bruxelles, 
\\50 Avenue F. D. Roosevelt, CP 194/5, B-1050 Bruxelles, Belgium
\\
$^4$Dipartimento di Ingegneria dell'Informazione, Universit\`a di Brescia, 
\\via Branze 38, I-25123 Brescia, Italy
\\
$^5$The Dodd-Walls Centre for Photonic and Quantum Technologies, Department of Physics, 
\\The University of Auckland, Auckland 1142, New Zealand
\\
$^6$CNR-INO, Istituto Nazionale di Ottica, Largo E. Fermi 6, I-50125 Firenze, Italy
\\
$^7$Dipartimento di Ingegneria dell'Informazione, Universit\`a di Brescia, 
\\and CNR-INO, via Branze 38, I-25123 Brescia
\\
$^8$
Novosibirsk State University, 1 Pirogova street, Novosibirsk 630090, Russia
}
\end{center}
\vskip 33pt

\noindent
{\large \bf 1. Mean field equation of a doubly resonant degenerate OPO}
\vskip 3pt
\noindent
Field dynamics of our OPO system can be described by an infinite dimensional map for the field amplitudes~\cite{Leo:2016kj,Leo:2016df}. First, propagation of the cavity fields over the $m$th round trip is governed by the amplitude evolution equations for the fields $A_m(z,\tau)$, at $\omega_0$, and $B_m(z,\tau)$, at $2\omega_0$:
\begin{align}
\frac{\partial A_m}{\partial z} =& \left[-\frac{\alpha_{c1}}{2}- i\frac{{k}_1''}{2}\frac{\partial^2}{\partial \tau^2}\right] \hspace{-3pt} A_m+i\kappa B_mA_m^*e^{-i \Delta k z}, 
\label{mapA}\\
\frac{\partial B_m}{\partial z} = &\left[-\frac{\alpha_{c2}}{2} - \Delta {k}'\frac{\partial }{\partial \tau}-i\frac{{k}_2''}{2}\frac{\partial^2 }{\partial \tau^2}\right] \hspace{-3pt} B_m+i\kappa A_m^2 e^{i\Delta k z} \, ,
\label{mapB}
\end{align}
where 
$z\in [0,L]$ is the position along the cavity round-trip path;
$\alpha_{c1,2}$ are propagation losses (hereafter, subscripts 1 and 2 denote fields at $\omega_0$ and $2\omega_0$, respectively); 
${k}''_{1,2} = \mathrm{d}^2k/\mathrm{d}\omega^2|_{\omega_0, 2\omega_0}$ are the group velocity dispersion coefficients; 
$\Delta {k}' = \mathrm{d}k/\mathrm{d}\omega|_{2\omega_0}-\mathrm{d}k/\mathrm{d}\omega|_{\omega_0}$ is the corresponding group-velocity mismatch, or temporal walk-off; 
$\Delta k=2 k(\omega_0)- k(2\omega_0)$ is the wave vector mismatch at the degeneracy.
 The nonlinear coupling constant $\kappa=\sqrt{8}\omega_0 \chi^{(2)}_\mathrm{eff} /\sqrt{c^3 n_1^2  n_2 \epsilon_0}$ is normalized such that $|A|^2$, $|B|^2$, and $|B_\mathrm{in}|^2$ are measured in watts, where $\chi^{(2)}_\mathrm{eff}$ is the effective second-order susceptibility, $c$ is the speed of light, $n_{1,2}$ are the refractive indices, and $\epsilon_0$ is the vacuum permittivity.
 The “fast-time” variable $\tau$ describes the temporal profiles of the fields in a reference frame moving with the group velocity of light at $\omega_0$. 
Second, the fields at the beginning of the $(m+1)$th round trip are related to the fields at the end of the $m$th one,
\begin{align}
A_{m+1}(0,\tau) &= \sqrt{1-\theta_1} \, A_{m}(L,\tau) \, e^{-i\delta_1}
\label{boundA} 
\\
B_{m+1}(0,\tau) &= B_\mathrm{in}  ,
\label{boundB} 
\end{align}
where $\theta_1$ is the power transmission of the coupling mirror at frequency $\omega_0$; 
$\delta_{1}\simeq  (\omega_0- \omega_c) t_\mathrm{R}$ is the round-trip phase detuning.

Following the approach described in Refs.~\cite{Leo:2016kj,Hansson:2017cs}, we first solve Eq.~(\ref{mapB}) in Fourier space:
\begin{align}
\mathscr{F} \left[B_m(z,\tau)\right] \approx 
& 
\mathscr{F}\left[B_m(0,\tau)\right] e^{\hat{k}z}
 + \kappa \mathscr{F}\left[A_m^2(z,\tau)\right] \displaystyle\frac{e^{i\Delta k z}-e^{\hat{k}z}}{\Delta k+i\hat{k}},
\label{slaved_SH}
\end{align}
where $\hat{k} \equiv \hat{k}(\Omega)  = -\alpha_{c,2}/2+i\left[\Delta{k}'\Omega+({k}_2''/2)\Omega^2\right]$, and
$\mathscr{F}\left[\cdot\right]=\int_{-\infty}^{\infty}\cdot\,e^{i\Omega\tau}\,\mathrm{d}\tau$ is the Fourier transform operator.
Writing explicitely
\[\mathscr{F}\left[B_m(0,\tau)\right] = \int_{-\infty}^{\infty} B_\mathrm{in}\,e^{i\Omega\tau}\,\mathrm{d}\tau  = B_\mathrm{in} \, \delta(\Omega) \, ,
\]
where $\delta(\Omega)$ is the Dirac function, we can transform back to time domain, obtaining
\begin{align}
B_m(z,\tau) = & B_\mathrm{in} \, e^{ -\frac{\alpha_{c2} z}{2} } 
+ \kappa \hspace{-2pt} \int_{-\infty}^{\infty} \hspace{-5pt} \mathscr{F}\left[A_m^2(z,\tau)\right]  \displaystyle\frac{e^{i\Delta k z}-e^{\hat{k}z}}{\Delta k+i\hat{k}} \,e^{-i\Omega\tau}\,\mathrm{d}\Omega \, .
 \label{EqA2}
\end{align}
We neglect SH linear losses (i.e., $\alpha_{c2}=0$, thus $\mathrm{Im}[\hat{I}(0)] = 0$), substitute the last expression into Eq.~(\ref{mapA}),  and integrate over one round trip, assuming $A_m(z,\tau)$ constant and keeping the leading order terms. This procedure eventually gives us the mean field Eq. (1) of the main text.

It is straightforward to demonstrate that the mean-field equation has a  constant (time independent) solution $A_0(t,\tau)=|A_0| \, e^{i \phi}$, which for $|A_0|=0$ is the trivial solution of the sub-threshold OPO.
We calculate the explicit expressions of $|A_0|$ and $\phi$ of the constant solution, as a function of the OPO parameters, assuming the phase-matching condition $\xi=0$.
We substitute the constant solution in the mean-field equation (Eq.~(1) in the main text) and obtain the following equation
\begin{equation}
-(\alpha_1 + i \delta_1) e^{2 i \phi} - \mu^2 \hat{I}(0) |A_0|^2 e^{2 i \phi} + i \mu  B_\mathrm{in}  =0 \, ,
\label{EqHom1}
\end{equation}
which can be written as a set of two equations for the real and imaginary parts,
\begin{align}
-\alpha_1 \cos 2\phi +\delta_1 \sin 2\phi -\mu^2 \hat{I}(0) |A_0|^2 \cos 2\phi &= 0 \\
\alpha_1 \sin 2\phi +\delta_1 \cos 2\phi  + \mu^2 \hat{I}(0) |A_0|^2 \sin 2\phi -\mu B_\mathrm{in} &= 0 \, ,
\end{align}
with  solutions 
\begin{equation}
\cos2\phi=\frac{\delta_1}{\mu B_\mathrm{in} } 
, \quad\quad
|A_0|^2=\frac{-\alpha_1 \pm \sqrt{\mu^2 B_\mathrm{in}^2 -\delta_1^2}  }{\mu^2  \, \hat{I}(0)} \, .
\label{EqHom}
\end{equation}
The physically meaningful solution, for which $|A_0|^2>0$, is with the higher sign, provided that $B_\mathrm{in}^2 > \eta_\mathrm{OPO} \equiv (\alpha_1^2 + \delta_1^2)/\mu^2$, where the last expression is the usual threshold expression for the dispersionless OPO.

Numerical simulation results have been obtained by integrating Eq.~(1) of the main text over the slow time $t$ by a split-step Fourier method, with a 4th-order Runge--Kutta scheme for evaluation of the nonlinear term.  
The buildup of the comb is seeded by random initial conditions, which mimic vacuum fluctuations. 
We numerically checked that the full map of Eqs.~(\ref{mapA})-(\ref{boundB}) and the mean-field equation yield almost identical results, though at the cost, for the full map, of a longer computation time.

\vskip 24pt
\noindent
{\large \bf 2. Modulation instability analysis}
\vskip 3pt
\noindent
Let us analyze the stability of the constant solution against the growth of new frequency components through modulation instability (MI). 
To this purpose, we introduce in Eq.~(1) of the main text the ansatz $A(t,\tau) = A_0 + a_1 e^{i \Omega \tau} + a_2 e^{ - i \Omega \tau}$, retaining terms up to the first order in the perturbations $a_1$ and $a_2$. Projecting onto each frequency component we get the linearized differential equations for the three field amplitudes:
\begin{subequations}
\begin{align}
\dot{A}_0 =&
 - ( \alpha_1 + i  \delta_1 ) \, A_0 
 - \mu^2 |A_0|^2 A_0 \, \hat{I}(0) 
  + i \mu B_\mathrm{in} A_0^*  \, ,
 \label{cav1} 
 \\
\dot{a}_1 =& 
- \left[ \alpha_1 + i \left( \delta_1 - D_2 \Omega^2 \right) + 2 \mu^2 |A_0|^2 \hat{I}(-\Omega)   \right]  a_1 
- [ \mu^2 A_0^2 \hat{I}(0) - i \mu B_\mathrm{in}  ]  \, a_2^* \, , 
\label{cav2} 
 \\
\dot{a}_2 =& 
- \left[ \alpha_1 + i \left( \delta_1 - D_2 \Omega^2 \right) + 2 \mu^2 |A_0|^2 \hat{I}(\Omega)   \right]  a_2
  - [ \mu^2 A_0^2 \hat{I}(0) - i \mu B_\mathrm{in} ]  \,  a_1^* \, ,
 \label{cav3} 
\end{align}
\label{cav} 
\end{subequations}

\noindent
where $D_2= L {k}_1''/2 $.
Eq.~(\ref{cav2}) and (\ref{cav3}) can be written in matrix form,
\begin{equation}
\left(\begin{array}{c} \dot{a}_1^* \\ \dot{a}_2 \end{array}\right)
=
M
\left(\begin{array}{c} a_1^* \\ a_2 \end{array}\right) \, ,
\end{equation}
where the elements of the matrix $M$ are
\begin{align*}
M_{11} =&  - \left[ \alpha_1 - i \left( \delta_1 - D_2 \Omega^2 \right) + 2 \mu^2 |A_0|^2 \hat{I}^*(-\Omega)   \right] ,
\\
M_{22} =& - \left[ \alpha_1 + i \left( \delta_1 - D_2 \Omega^2 \right) + 2 \mu^2 |A_0|^2 \hat{I}(\Omega)   \right] ,
\\
M_{12}^* =& M_{21} =   - \mu^2 A_0^2 \hat{I}(0) + i \mu B_\mathrm{in} 
= (\alpha_1 + \delta_1) \, e^{2 i\phi} \, ,
\end{align*}
the last equality being derived from Eq.~(\ref{EqHom1}) (this equality does not hold for the trivial zero solution).  
The eigenvalues of the matrix $M$ are
\begin{equation}
\lambda_{\pm} =  
- \left[ \alpha_1 + \mu^2 |A_0|^2 {\cal I}_+(\Omega) \right] 
\nonumber
 \pm
\sqrt{ 
(\alpha_1^2 + \delta_1^2) \hspace{-2pt}
- \hspace{-2pt} \left[ \delta_1 \hspace{-2pt} - \hspace{-2pt}D_2  \Omega^2  - i \mu^2 |A_0|^2 \, {\cal I}_-(\Omega)  \right]^2
} \, ,
\end{equation}
where $|A_0|$ depends on $B_\mathrm{in}$ through Eqs.~(\ref{EqHom}) and ${\cal I}_\pm(\Omega) = \hat{I}(\Omega) \pm \hat{I}^*(-\Omega)$. 
The nontrivial constant solution becomes unstable when the MI gain $\mathrm{Re}[\lambda_+]>0$.

\begin{figure}[t!]
\begin{center}
\includegraphics*[viewport= 0 0 580 290, clip, width=0.8\columnwidth]{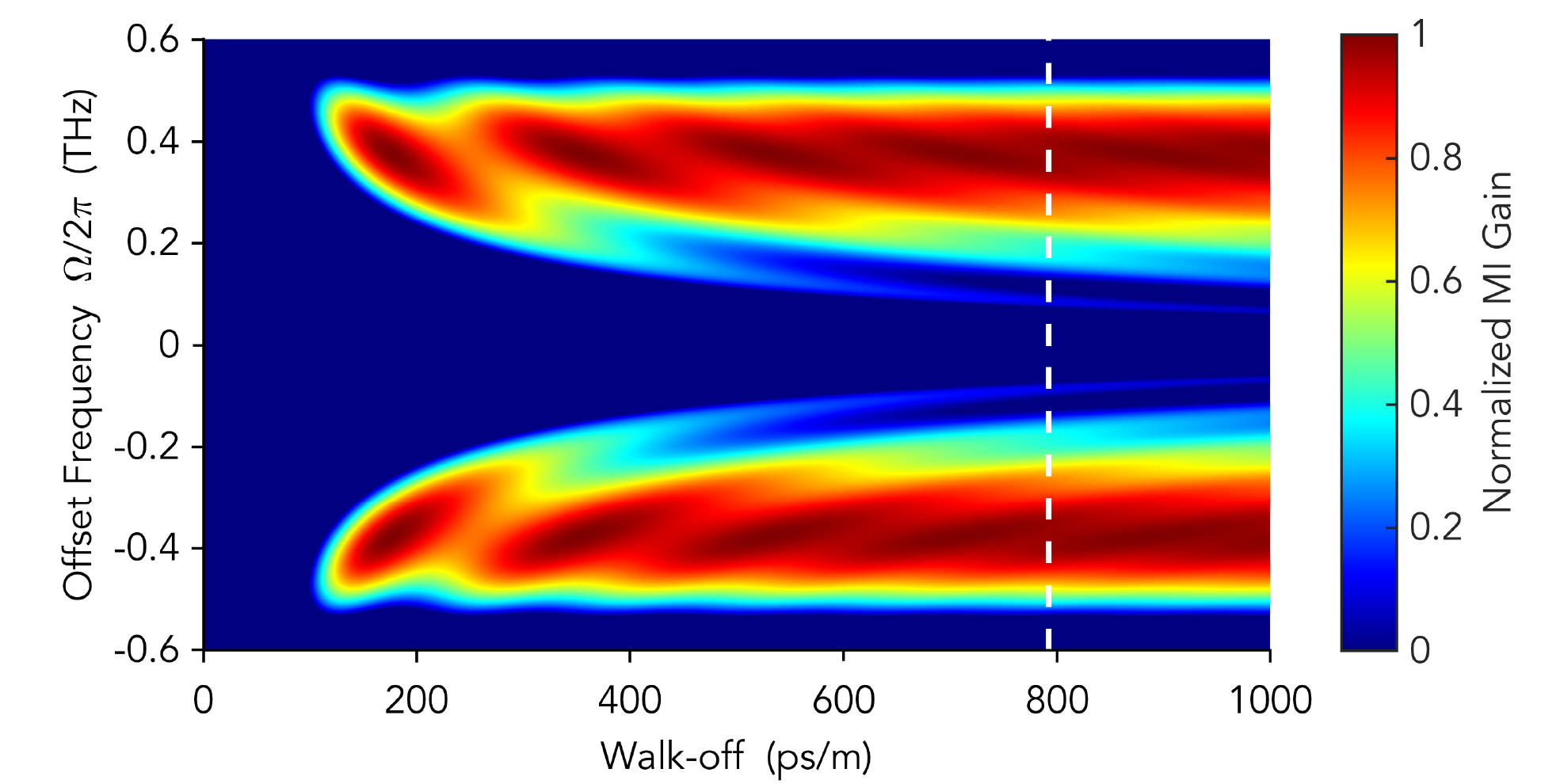}
\caption{ 
MI gain as a function of the walk-off parameter $\Delta  k^\prime$, for $\delta_1=0.01$ ($\Delta \simeq 0.5$). The dashed line marks the walk-off parameter of our experimental cavity, $\Delta  k^\prime=792$~ps/m.}
\label{Fig7}
\end{center}
\end{figure}

Analogously,  the trivial zero solution exhibits MI gain for positive real part of the eigenvalues 
\begin{align}
\lambda_{\pm} =&  
-  \alpha_1 \pm
\sqrt{ \mu^2 B_\mathrm{in}^2
- \left( \delta_1 - D_2  \Omega^2 \right)^2} \, .
\end{align}


MI gain for the constant finite and zero solutions is displayed in Fig.~1 of the main text. 
Figure~\ref{Fig7} shows the MI gain for the nontrivial solution as a  function of the offset frequency and of temporal walk-off, for $\delta_1=0.01$, ($\Delta \simeq0.5$).

\vskip 24pt
\noindent
{\large \bf 3. Experimental setup}
\vskip 3pt
\noindent
 The OPO cavity consists of four mirrors in a bow-tie configuration. 
The periodically-poled 5\%-MgO-doped lithium niobate crystal is located between two high-reflectivity (HR) ($R>99.9\%$) spherical mirrors (radius of curvature, 100~mm).
A flat HR mirror is mounted on a piezoelectric actuator (PZT) for cavity length control. A fourth, partially reflective flat mirror (R= 98\%)  allows to couple out the generated parametric radiation.  The crystal temperature is actively stabilized, within 1~mK, by a Peltier element driven by an electronic servo control. 
All the mirrors have an anti-reflective coating for the SH green light.

The spectral composition of the parametric waves is analyzed by an optical spectrum analyzer with a detection range from 600 to 1700~nm and a resolution bandwidth $\mathrm{RBW}=6~\mathrm{GHz}$.  
A few mW of the parametric beam is also sent to a fast p-i-n InGaAs photodiode (response bandwidth 2~GHz), whose ac signal is processed by a radio frequency (RF) spectrum analyzer. 
Similarly,  the SH light leaving the OPO cavity is analyzed by a fast Si photodiode (response bandwidth 2 GHz) and a second RF spectrum analyzer. 
In their respective optical ranges, the two photodiode RF spectra provide evidence of comb line spacing on the scale of a few FSR with a resolution of 1~kHz.

\begin{figure}[t!]
\begin{center}
\includegraphics*[viewport= 0 0 550 340, clip, width=0.8\columnwidth]{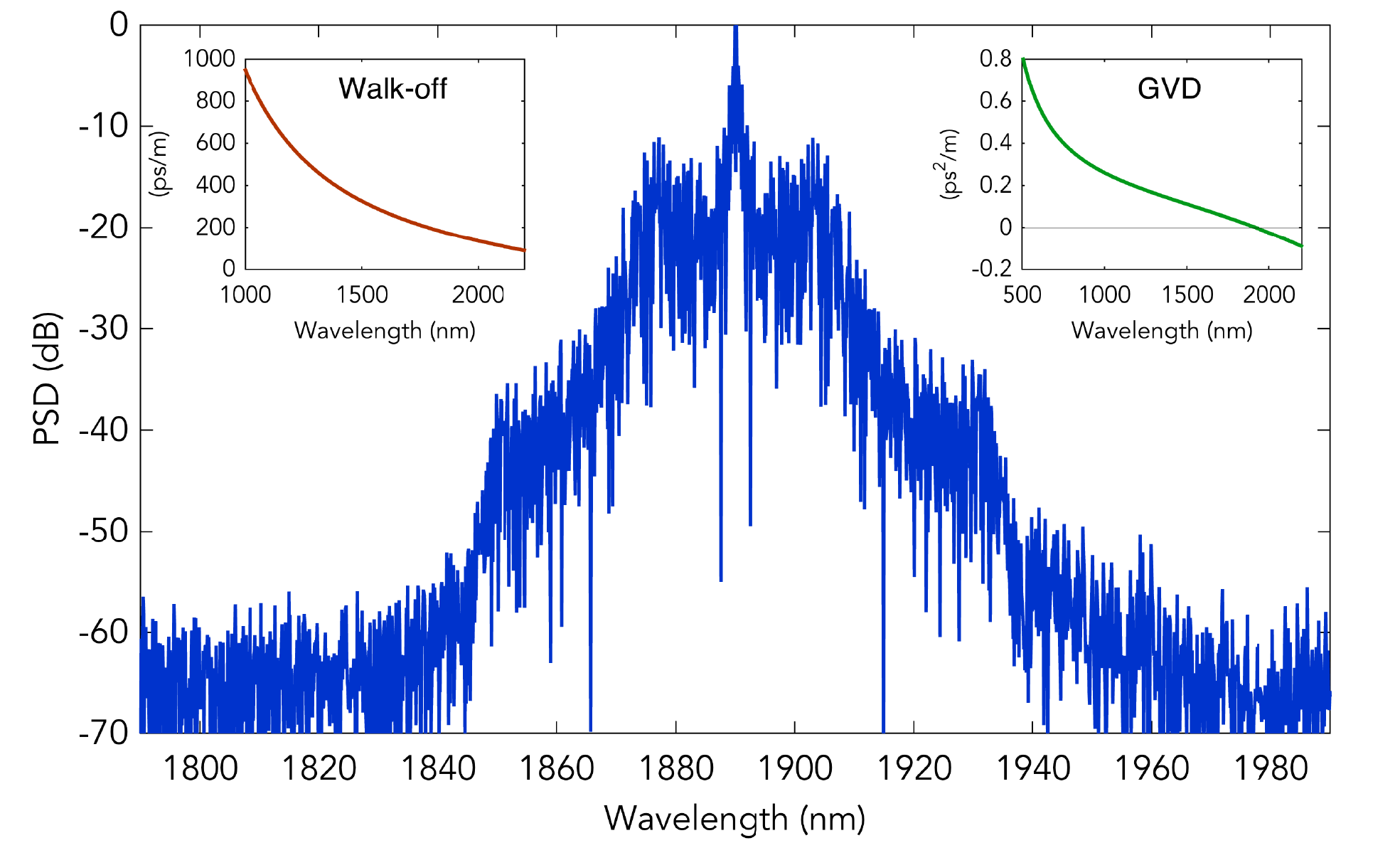}
\caption{ 
Effect of dispersion on comb frequency span. 
Spectrum calculated for a degenerate OPO pumped at 945~nm, with signal field at 1890~nm, a wavelength for which the group velocity dispersion coefficient of the material reduces ($P_\mathrm{in}=300$~mW, $\Delta=0$). The insets show walk-off parameter and group velocity dispersion coefficient, respectively, as a function of the wavelength. }
\label{Fig8}
\end{center}
\end{figure}

\vskip 24pt
\noindent
{\large \bf 4. Effect of dispersion on comb frequency span}
\vskip 3pt
\noindent
Figure~\ref{Fig8} shows a comb spectrum calculated for a degenerate OPO with signal field at 1890~nm, where group velocity dispersion strongly reduces with respect to 1064~nm (see the inset on the right). In this case, dispersion has been included up to the third order. 
In particular, the walk-off parameter is $\Delta  k^\prime=166$~ps/m, ${k}''_{1890} = 8.0 \times 10^{-3}\;\mathrm{ps^2/m}$,  ${k}''_{945} =0.284\;\mathrm{ps^2/m}$, ${k}'''_{1890} = 5.69 \times 10^{-4}\;\mathrm{ps^3/m}$,  ${k}'''_{945} =2.14 \times 10^{-4}\;\mathrm{ps^3/m}$.

\begin{thebibliography}{59}%
\makeatletter
\providecommand \@ifxundefined [1]{%
 \@ifx{#1\undefined}
}%
\providecommand \@ifnum [1]{%
 \ifnum #1\expandafter \@firstoftwo
 \else \expandafter \@secondoftwo
 \fi
}%
\providecommand \@ifx [1]{%
 \ifx #1\expandafter \@firstoftwo
 \else \expandafter \@secondoftwo
 \fi
}%
\providecommand \natexlab [1]{#1}%
\providecommand \enquote  [1]{``#1''}%
\providecommand \bibnamefont  [1]{#1}%
\providecommand \bibfnamefont [1]{#1}%
\providecommand \citenamefont [1]{#1}%
\providecommand \href@noop [0]{\@secondoftwo}%
\providecommand \href [0]{\begingroup \@sanitize@url \@href}%
\providecommand \@href[1]{\@@startlink{#1}\@@href}%
\providecommand \@@href[1]{\endgroup#1\@@endlink}%
\providecommand \@sanitize@url [0]{\catcode `\\12\catcode `\$12\catcode
  `\&12\catcode `\#12\catcode `\^12\catcode `\_12\catcode `\%12\relax}%
\providecommand \@@startlink[1]{}%
\providecommand \@@endlink[0]{}%
\providecommand \url  [0]{\begingroup\@sanitize@url \@url }%
\providecommand \@url [1]{\endgroup\@href {#1}{\urlprefix }}%
\providecommand \urlprefix  [0]{URL }%
\providecommand \Eprint [0]{\href }%
\providecommand \doibase [0]{http://dx.doi.org/}%
\providecommand \selectlanguage [0]{\@gobble}%
\providecommand \bibinfo  [0]{\@secondoftwo}%
\providecommand \bibfield  [0]{\@secondoftwo}%
\providecommand \translation [1]{[#1]}%
\providecommand \BibitemOpen [0]{}%
\providecommand \bibitemStop [0]{}%
\providecommand \bibitemNoStop [0]{.\EOS\space}%
\providecommand \EOS [0]{\spacefactor3000\relax}%
\providecommand \BibitemShut  [1]{\csname bibitem#1\endcsname}%
\let\auto@bib@innerbib\@empty
\bibitem [{\citenamefont {Holzwarth}\ \emph {et~al.}(2000)\citenamefont
  {Holzwarth}, \citenamefont {Udem}, \citenamefont {H\"ansch}, \citenamefont
  {Knight}, \citenamefont {Wadsworth},\ and\ \citenamefont
  {Russell}}]{Holzwarth:2000aa}%
  \BibitemOpen
  \bibfield  {author} {\bibinfo {author} {\bibfnamefont {R.}~\bibnamefont
  {Holzwarth}}, \bibinfo {author} {\bibfnamefont {T.}~\bibnamefont {Udem}},
  \bibinfo {author} {\bibfnamefont {T.~W.}\ \bibnamefont {H\"ansch}}, \bibinfo
  {author} {\bibfnamefont {J.~C.}\ \bibnamefont {Knight}}, \bibinfo {author}
  {\bibfnamefont {W.~J.}\ \bibnamefont {Wadsworth}}, \ and\ \bibinfo {author}
  {\bibfnamefont {P.~S.~J.}\ \bibnamefont {Russell}},\ }\href@noop {}
  {\bibfield  {journal} {\bibinfo  {journal} {Phys. Rev. Lett.}\ }\textbf
  {\bibinfo {volume} {85}},\ \bibinfo {pages} {2264} (\bibinfo {year}
  {2000})}\BibitemShut {NoStop}%
\bibitem [{\citenamefont {Jones}\ \emph {et~al.}(2000)\citenamefont {Jones},
  \citenamefont {Diddams}, \citenamefont {Ranka}, \citenamefont {Stentz},
  \citenamefont {Windeler}, \citenamefont {Hall},\ and\ \citenamefont
  {Cundiff}}]{Jones:2000tn}%
  \BibitemOpen
  \bibfield  {author} {\bibinfo {author} {\bibfnamefont {D.~J.}\ \bibnamefont
  {Jones}}, \bibinfo {author} {\bibfnamefont {S.~A.}\ \bibnamefont {Diddams}},
  \bibinfo {author} {\bibfnamefont {J.~K.}\ \bibnamefont {Ranka}}, \bibinfo
  {author} {\bibfnamefont {A.~J.}\ \bibnamefont {Stentz}}, \bibinfo {author}
  {\bibfnamefont {R.~S.}\ \bibnamefont {Windeler}}, \bibinfo {author}
  {\bibfnamefont {J.~L.}\ \bibnamefont {Hall}}, \ and\ \bibinfo {author}
  {\bibfnamefont {S.~T.}\ \bibnamefont {Cundiff}},\ }\href@noop {} {\bibfield
  {journal} {\bibinfo  {journal} {Science}\ }\textbf {\bibinfo {volume}
  {288}},\ \bibinfo {pages} {635} (\bibinfo {year} {2000})}\BibitemShut
  {NoStop}%
\bibitem [{\citenamefont {Del'Haye}\ \emph {et~al.}(2007)\citenamefont
  {Del'Haye}, \citenamefont {Schliesser}, \citenamefont {Arcizet},
  \citenamefont {Wilken}, \citenamefont {Holzwarth},\ and\ \citenamefont
  {Kippenberg}}]{DelHaye:2007gi}%
  \BibitemOpen
  \bibfield  {author} {\bibinfo {author} {\bibfnamefont {P.}~\bibnamefont
  {Del'Haye}}, \bibinfo {author} {\bibfnamefont {A.}~\bibnamefont
  {Schliesser}}, \bibinfo {author} {\bibfnamefont {O.}~\bibnamefont {Arcizet}},
  \bibinfo {author} {\bibfnamefont {T.}~\bibnamefont {Wilken}}, \bibinfo
  {author} {\bibfnamefont {R.}~\bibnamefont {Holzwarth}}, \ and\ \bibinfo
  {author} {\bibfnamefont {T.~J.}\ \bibnamefont {Kippenberg}},\ }\href@noop {}
  {\bibfield  {journal} {\bibinfo  {journal} {Nature}\ }\textbf {\bibinfo
  {volume} {450}},\ \bibinfo {pages} {1214} (\bibinfo {year}
  {2007})}\BibitemShut {NoStop}%
\bibitem [{\citenamefont {Chembo}\ \emph {et~al.}(2010)\citenamefont {Chembo},
  \citenamefont {Strekalov},\ and\ \citenamefont {Yu}}]{Chembo:2010ii}%
  \BibitemOpen
  \bibfield  {author} {\bibinfo {author} {\bibfnamefont {Y.~K.}\ \bibnamefont
  {Chembo}}, \bibinfo {author} {\bibfnamefont {D.~V.}\ \bibnamefont
  {Strekalov}}, \ and\ \bibinfo {author} {\bibfnamefont {N.}~\bibnamefont
  {Yu}},\ }\href@noop {} {\bibfield  {journal} {\bibinfo  {journal} {Phys. Rev.
  Lett.}\ }\textbf {\bibinfo {volume} {104}},\ \bibinfo {pages} {103902}
  (\bibinfo {year} {2010})}\BibitemShut {NoStop}%
\bibitem [{\citenamefont {Hansson}\ \emph {et~al.}(2013)\citenamefont
  {Hansson}, \citenamefont {Modotto},\ and\ \citenamefont
  {Wabnitz}}]{Hansson:2013jy}%
  \BibitemOpen
  \bibfield  {author} {\bibinfo {author} {\bibfnamefont {T.}~\bibnamefont
  {Hansson}}, \bibinfo {author} {\bibfnamefont {D.}~\bibnamefont {Modotto}}, \
  and\ \bibinfo {author} {\bibfnamefont {S.}~\bibnamefont {Wabnitz}},\
  }\href@noop {} {\bibfield  {journal} {\bibinfo  {journal} {Phys. Rev. A}\
  }\textbf {\bibinfo {volume} {88}},\ \bibinfo {pages} {023819} (\bibinfo
  {year} {2013})}\BibitemShut {NoStop}%
\bibitem [{\citenamefont {Kippenberg}\ \emph {et~al.}(2011)\citenamefont
  {Kippenberg}, \citenamefont {Holzwarth},\ and\ \citenamefont
  {Diddams}}]{Kippenberg:2011fc}%
  \BibitemOpen
  \bibfield  {author} {\bibinfo {author} {\bibfnamefont {T.~J.}\ \bibnamefont
  {Kippenberg}}, \bibinfo {author} {\bibfnamefont {R.}~\bibnamefont
  {Holzwarth}}, \ and\ \bibinfo {author} {\bibfnamefont {S.~A.}\ \bibnamefont
  {Diddams}},\ }\href@noop {} {\bibfield  {journal} {\bibinfo  {journal}
  {Science}\ }\textbf {\bibinfo {volume} {332}},\ \bibinfo {pages} {555}
  (\bibinfo {year} {2011})}\BibitemShut {NoStop}%
\bibitem [{\citenamefont {Ulvila}\ \emph {et~al.}(2013)\citenamefont {Ulvila},
  \citenamefont {Phillips}, \citenamefont {Halonen},\ and\ \citenamefont
  {Vainio}}]{Ulvila:2013jv}%
  \BibitemOpen
  \bibfield  {author} {\bibinfo {author} {\bibfnamefont {V.}~\bibnamefont
  {Ulvila}}, \bibinfo {author} {\bibfnamefont {C.~R.}\ \bibnamefont
  {Phillips}}, \bibinfo {author} {\bibfnamefont {L.~L.}\ \bibnamefont
  {Halonen}}, \ and\ \bibinfo {author} {\bibfnamefont {M.}~\bibnamefont
  {Vainio}},\ }\href@noop {} {\bibfield  {journal} {\bibinfo  {journal} {Opt.
  Lett.}\ }\textbf {\bibinfo {volume} {38}},\ \bibinfo {pages} {4281} (\bibinfo
  {year} {2013})}\BibitemShut {NoStop}%
\bibitem [{\citenamefont {Ulvila}\ \emph {et~al.}(2014)\citenamefont {Ulvila},
  \citenamefont {Phillips}, \citenamefont {Halonen},\ and\ \citenamefont
  {Vainio}}]{Ulvila:2014bx}%
  \BibitemOpen
  \bibfield  {author} {\bibinfo {author} {\bibfnamefont {V.}~\bibnamefont
  {Ulvila}}, \bibinfo {author} {\bibfnamefont {C.~R.}\ \bibnamefont
  {Phillips}}, \bibinfo {author} {\bibfnamefont {L.~L.}\ \bibnamefont
  {Halonen}}, \ and\ \bibinfo {author} {\bibfnamefont {M.}~\bibnamefont
  {Vainio}},\ }\href@noop {} {\bibfield  {journal} {\bibinfo  {journal} {Opt.
  Express}\ }\textbf {\bibinfo {volume} {22}},\ \bibinfo {pages} {10535}
  (\bibinfo {year} {2014})}\BibitemShut {NoStop}%
\bibitem [{\citenamefont {Ricciardi}\ \emph {et~al.}(2015)\citenamefont
  {Ricciardi}, \citenamefont {Mosca}, \citenamefont {Parisi}, \citenamefont
  {Maddaloni}, \citenamefont {Santamaria}, \citenamefont {De~Natale},\ and\
  \citenamefont {De~Rosa}}]{Ricciardi:2015bw}%
  \BibitemOpen
  \bibfield  {author} {\bibinfo {author} {\bibfnamefont {I.}~\bibnamefont
  {Ricciardi}}, \bibinfo {author} {\bibfnamefont {S.}~\bibnamefont {Mosca}},
  \bibinfo {author} {\bibfnamefont {M.}~\bibnamefont {Parisi}}, \bibinfo
  {author} {\bibfnamefont {P.}~\bibnamefont {Maddaloni}}, \bibinfo {author}
  {\bibfnamefont {L.}~\bibnamefont {Santamaria}}, \bibinfo {author}
  {\bibfnamefont {P.}~\bibnamefont {De~Natale}}, \ and\ \bibinfo {author}
  {\bibfnamefont {M.}~\bibnamefont {De~Rosa}},\ }\href@noop {} {\bibfield
  {journal} {\bibinfo  {journal} {Phys. Rev. A}\ }\textbf {\bibinfo {volume}
  {91}},\ \bibinfo {pages} {063839} (\bibinfo {year} {2015})}\BibitemShut
  {NoStop}%
\bibitem [{\citenamefont {Mosca}\ \emph {et~al.}(2016)\citenamefont {Mosca},
  \citenamefont {Ricciardi}, \citenamefont {Parisi}, \citenamefont {Maddaloni},
  \citenamefont {Santamaria}, \citenamefont {De~Natale},\ and\ \citenamefont
  {De~Rosa}}]{Mosca:2015wh}%
  \BibitemOpen
  \bibfield  {author} {\bibinfo {author} {\bibfnamefont {S.}~\bibnamefont
  {Mosca}}, \bibinfo {author} {\bibfnamefont {I.}~\bibnamefont {Ricciardi}},
  \bibinfo {author} {\bibfnamefont {M.}~\bibnamefont {Parisi}}, \bibinfo
  {author} {\bibfnamefont {P.}~\bibnamefont {Maddaloni}}, \bibinfo {author}
  {\bibfnamefont {L.}~\bibnamefont {Santamaria}}, \bibinfo {author}
  {\bibfnamefont {P.}~\bibnamefont {De~Natale}}, \ and\ \bibinfo {author}
  {\bibfnamefont {M.}~\bibnamefont {De~Rosa}},\ }\href@noop {} {\bibfield
  {journal} {\bibinfo  {journal} {Nanophotonics}\ }\textbf {\bibinfo {volume}
  {5}},\ \bibinfo {pages} {316} (\bibinfo {year} {2016})}\BibitemShut {NoStop}%
\bibitem [{\citenamefont {Schiller}\ \emph {et~al.}(1996)\citenamefont
  {Schiller}, \citenamefont {Breitenbach}, \citenamefont {Paschotta},\ and\
  \citenamefont {Mlynek}}]{Schiller:1996gx}%
  \BibitemOpen
  \bibfield  {author} {\bibinfo {author} {\bibfnamefont {S.}~\bibnamefont
  {Schiller}}, \bibinfo {author} {\bibfnamefont {G.}~\bibnamefont
  {Breitenbach}}, \bibinfo {author} {\bibfnamefont {R.~R.}\ \bibnamefont
  {Paschotta}}, \ and\ \bibinfo {author} {\bibfnamefont {J.}~\bibnamefont
  {Mlynek}},\ }\href@noop {} {\bibfield  {journal} {\bibinfo  {journal} {Appl.
  Phys. Lett.}\ }\textbf {\bibinfo {volume} {68}},\ \bibinfo {pages} {3374}
  (\bibinfo {year} {1996})}\BibitemShut {NoStop}%
\bibitem [{\citenamefont {Schiller}\ \emph {et~al.}(1997)\citenamefont
  {Schiller}, \citenamefont {Bruckmeier},\ and\ \citenamefont
  {White}}]{Schiller:1997dp}%
  \BibitemOpen
  \bibfield  {author} {\bibinfo {author} {\bibfnamefont {S.}~\bibnamefont
  {Schiller}}, \bibinfo {author} {\bibfnamefont {R.}~\bibnamefont
  {Bruckmeier}}, \ and\ \bibinfo {author} {\bibfnamefont {A.~G.}\ \bibnamefont
  {White}},\ }\href@noop {} {\bibfield  {journal} {\bibinfo  {journal} {Opt.
  Commun.}\ }\textbf {\bibinfo {volume} {138}},\ \bibinfo {pages} {158}
  (\bibinfo {year} {1997})}\BibitemShut {NoStop}%
\bibitem [{\citenamefont {White}\ \emph {et~al.}(1997)\citenamefont {White},
  \citenamefont {Lam}, \citenamefont {Taubman}, \citenamefont {Marte},
  \citenamefont {Schiller}, \citenamefont {McClelland},\ and\ \citenamefont
  {Bachor}}]{White:1997ta}%
  \BibitemOpen
  \bibfield  {author} {\bibinfo {author} {\bibfnamefont {A.~G.}\ \bibnamefont
  {White}}, \bibinfo {author} {\bibfnamefont {P.~K.}\ \bibnamefont {Lam}},
  \bibinfo {author} {\bibfnamefont {M.~S.}\ \bibnamefont {Taubman}}, \bibinfo
  {author} {\bibfnamefont {M.~A.~M.}\ \bibnamefont {Marte}}, \bibinfo {author}
  {\bibfnamefont {S.}~\bibnamefont {Schiller}}, \bibinfo {author}
  {\bibfnamefont {D.~E.}\ \bibnamefont {McClelland}}, \ and\ \bibinfo {author}
  {\bibfnamefont {H.-A.}\ \bibnamefont {Bachor}},\ }\href@noop {} {\bibfield
  {journal} {\bibinfo  {journal} {Phys. Rev. A}\ }\textbf {\bibinfo {volume}
  {55}},\ \bibinfo {pages} {4511} (\bibinfo {year} {1997})}\BibitemShut
  {NoStop}%
\bibitem [{\citenamefont {Hansson}\ \emph {et~al.}(2016)\citenamefont
  {Hansson}, \citenamefont {Leo}, \citenamefont {Erkintalo}, \citenamefont
  {Anthony}, \citenamefont {Coen}, \citenamefont {Ricciardi}, \citenamefont
  {De~Rosa},\ and\ \citenamefont {Wabnitz}}]{Hansson:2016kz}%
  \BibitemOpen
  \bibfield  {author} {\bibinfo {author} {\bibfnamefont {T.}~\bibnamefont
  {Hansson}}, \bibinfo {author} {\bibfnamefont {F.}~\bibnamefont {Leo}},
  \bibinfo {author} {\bibfnamefont {M.}~\bibnamefont {Erkintalo}}, \bibinfo
  {author} {\bibfnamefont {J.}~\bibnamefont {Anthony}}, \bibinfo {author}
  {\bibfnamefont {S.}~\bibnamefont {Coen}}, \bibinfo {author} {\bibfnamefont
  {I.}~\bibnamefont {Ricciardi}}, \bibinfo {author} {\bibfnamefont
  {M.}~\bibnamefont {De~Rosa}}, \ and\ \bibinfo {author} {\bibfnamefont
  {S.}~\bibnamefont {Wabnitz}},\ }\href@noop {} {\bibfield  {journal} {\bibinfo
   {journal} {J. Opt. Soc. Am. B}\ }\textbf {\bibinfo {volume} {33}},\ \bibinfo
  {pages} {1207} (\bibinfo {year} {2016})}\BibitemShut {NoStop}%
\bibitem [{\citenamefont {Schliesser}\ \emph {et~al.}(2012)\citenamefont
  {Schliesser}, \citenamefont {Picqu{\'e}},\ and\ \citenamefont
  {H{\"a}nsch}}]{Schliesser:2012dn}%
  \BibitemOpen
  \bibfield  {author} {\bibinfo {author} {\bibfnamefont {A.}~\bibnamefont
  {Schliesser}}, \bibinfo {author} {\bibfnamefont {N.}~\bibnamefont
  {Picqu{\'e}}}, \ and\ \bibinfo {author} {\bibfnamefont {T.~W.}\ \bibnamefont
  {H{\"a}nsch}},\ }\href@noop {} {\bibfield  {journal} {\bibinfo  {journal}
  {Nature Photon.}\ }\textbf {\bibinfo {volume} {6}},\ \bibinfo {pages} {440}
  (\bibinfo {year} {2012})}\BibitemShut {NoStop}%
\bibitem [{\citenamefont {Maddaloni}\ \emph {et~al.}(2009)\citenamefont
  {Maddaloni}, \citenamefont {Cancio~Pastor},\ and\ \citenamefont
  {De~Natale}}]{Maddaloni:2009bg}%
  \BibitemOpen
  \bibfield  {author} {\bibinfo {author} {\bibfnamefont {P.}~\bibnamefont
  {Maddaloni}}, \bibinfo {author} {\bibfnamefont {P.}~\bibnamefont
  {Cancio~Pastor}}, \ and\ \bibinfo {author} {\bibfnamefont {P.}~\bibnamefont
  {De~Natale}},\ }\href@noop {} {\bibfield  {journal} {\bibinfo  {journal}
  {Meas. Sci. Technol.}\ }\textbf {\bibinfo {volume} {20}},\ \bibinfo {pages}
  {052001} (\bibinfo {year} {2009})}\BibitemShut {NoStop}%
\bibitem [{\citenamefont {Maddaloni}\ \emph {et~al.}(2006)\citenamefont
  {Maddaloni}, \citenamefont {Malara}, \citenamefont {Gagliardi},\ and\
  \citenamefont {De~Natale}}]{Maddaloni:2006ka}%
  \BibitemOpen
  \bibfield  {author} {\bibinfo {author} {\bibfnamefont {P.}~\bibnamefont
  {Maddaloni}}, \bibinfo {author} {\bibfnamefont {P.}~\bibnamefont {Malara}},
  \bibinfo {author} {\bibfnamefont {G.}~\bibnamefont {Gagliardi}}, \ and\
  \bibinfo {author} {\bibfnamefont {P.}~\bibnamefont {De~Natale}},\ }\href@noop
  {} {\bibfield  {journal} {\bibinfo  {journal} {New J. Phys.}\ }\textbf
  {\bibinfo {volume} {8}},\ \bibinfo {pages} {262} (\bibinfo {year}
  {2006})}\BibitemShut {NoStop}%
\bibitem [{\citenamefont {Erny}\ \emph {et~al.}(2007)\citenamefont {Erny},
  \citenamefont {Moutzouris}, \citenamefont {Biegert}, \citenamefont
  {K{\"u}hlke}, \citenamefont {Adler}, \citenamefont {Leitenstorfer},\ and\
  \citenamefont {Keller}}]{Erny:2007dj}%
  \BibitemOpen
  \bibfield  {author} {\bibinfo {author} {\bibfnamefont {C.}~\bibnamefont
  {Erny}}, \bibinfo {author} {\bibfnamefont {K.}~\bibnamefont {Moutzouris}},
  \bibinfo {author} {\bibfnamefont {J.}~\bibnamefont {Biegert}}, \bibinfo
  {author} {\bibfnamefont {D.}~\bibnamefont {K{\"u}hlke}}, \bibinfo {author}
  {\bibfnamefont {F.}~\bibnamefont {Adler}}, \bibinfo {author} {\bibfnamefont
  {A.}~\bibnamefont {Leitenstorfer}}, \ and\ \bibinfo {author} {\bibfnamefont
  {U.}~\bibnamefont {Keller}},\ }\href@noop {} {\bibfield  {journal} {\bibinfo
  {journal} {Opt. Lett.}\ }\textbf {\bibinfo {volume} {32}},\ \bibinfo {pages}
  {1138} (\bibinfo {year} {2007})}\BibitemShut {NoStop}%
\bibitem [{\citenamefont {Sun}\ \emph {et~al.}(2007)\citenamefont {Sun},
  \citenamefont {Gale},\ and\ \citenamefont {Reid}}]{Sun:2007wu}%
  \BibitemOpen
  \bibfield  {author} {\bibinfo {author} {\bibfnamefont {J.~H.}\ \bibnamefont
  {Sun}}, \bibinfo {author} {\bibfnamefont {B.~J.~S.}\ \bibnamefont {Gale}}, \
  and\ \bibinfo {author} {\bibfnamefont {D.~T.}\ \bibnamefont {Reid}},\
  }\href@noop {} {\bibfield  {journal} {\bibinfo  {journal} {Opt. Lett.}\
  }\textbf {\bibinfo {volume} {32}},\ \bibinfo {pages} {1414} (\bibinfo {year}
  {2007})}\BibitemShut {NoStop}%
\bibitem [{\citenamefont {Adler}\ \emph {et~al.}(2009)\citenamefont {Adler},
  \citenamefont {Cossel}, \citenamefont {Thorpe}, \citenamefont {Hartl},
  \citenamefont {Fermann},\ and\ \citenamefont {Ye}}]{Adler:2009ka}%
  \BibitemOpen
  \bibfield  {author} {\bibinfo {author} {\bibfnamefont {F.}~\bibnamefont
  {Adler}}, \bibinfo {author} {\bibfnamefont {K.~C.}\ \bibnamefont {Cossel}},
  \bibinfo {author} {\bibfnamefont {M.~J.}\ \bibnamefont {Thorpe}}, \bibinfo
  {author} {\bibfnamefont {I.}~\bibnamefont {Hartl}}, \bibinfo {author}
  {\bibfnamefont {M.~E.}\ \bibnamefont {Fermann}}, \ and\ \bibinfo {author}
  {\bibfnamefont {J.}~\bibnamefont {Ye}},\ }\href@noop {} {\bibfield  {journal}
  {\bibinfo  {journal} {Opt. Lett.}\ }\textbf {\bibinfo {volume} {34}},\
  \bibinfo {pages} {1330} (\bibinfo {year} {2009})}\BibitemShut {NoStop}%
\bibitem [{\citenamefont {Wong}\ \emph {et~al.}(2010)\citenamefont {Wong},
  \citenamefont {Vodopyanov},\ and\ \citenamefont {Byer}}]{Wong:2010gw}%
  \BibitemOpen
  \bibfield  {author} {\bibinfo {author} {\bibfnamefont {S.~T.}\ \bibnamefont
  {Wong}}, \bibinfo {author} {\bibfnamefont {K.~L.}\ \bibnamefont
  {Vodopyanov}}, \ and\ \bibinfo {author} {\bibfnamefont {R.~L.}\ \bibnamefont
  {Byer}},\ }\href@noop {} {\bibfield  {journal} {\bibinfo  {journal} {J. Opt.
  Soc. Am. B}\ }\textbf {\bibinfo {volume} {27}},\ \bibinfo {pages} {876}
  (\bibinfo {year} {2010})}\BibitemShut {NoStop}%
\bibitem [{\citenamefont {Leindecker}\ \emph {et~al.}(2011)\citenamefont
  {Leindecker}, \citenamefont {Marandi}, \citenamefont {Byer},\ and\
  \citenamefont {Vodopyanov}}]{Leindecker:2011ch}%
  \BibitemOpen
  \bibfield  {author} {\bibinfo {author} {\bibfnamefont {N.~C.}\ \bibnamefont
  {Leindecker}}, \bibinfo {author} {\bibfnamefont {A.}~\bibnamefont {Marandi}},
  \bibinfo {author} {\bibfnamefont {R.~L.}\ \bibnamefont {Byer}}, \ and\
  \bibinfo {author} {\bibfnamefont {K.~L.}\ \bibnamefont {Vodopyanov}},\
  }\href@noop {} {\bibfield  {journal} {\bibinfo  {journal} {Opt. Express}\
  }\textbf {\bibinfo {volume} {19}},\ \bibinfo {pages} {6296} (\bibinfo {year}
  {2011})}\BibitemShut {NoStop}%
\bibitem [{\citenamefont {Leindecker}\ \emph {et~al.}(2012)\citenamefont
  {Leindecker}, \citenamefont {Marandi}, \citenamefont {Byer}, \citenamefont
  {Vodopyanov}, \citenamefont {Jiang}, \citenamefont {Hartl}, \citenamefont
  {Fermann},\ and\ \citenamefont {Schunemann}}]{Leindecker:2012ed}%
  \BibitemOpen
  \bibfield  {author} {\bibinfo {author} {\bibfnamefont {N.~N.}\ \bibnamefont
  {Leindecker}}, \bibinfo {author} {\bibfnamefont {A.~A.}\ \bibnamefont
  {Marandi}}, \bibinfo {author} {\bibfnamefont {R.~L.~R.}\ \bibnamefont
  {Byer}}, \bibinfo {author} {\bibfnamefont {K.~L.}\ \bibnamefont
  {Vodopyanov}}, \bibinfo {author} {\bibfnamefont {J.~J.}\ \bibnamefont
  {Jiang}}, \bibinfo {author} {\bibfnamefont {I.~I.}\ \bibnamefont {Hartl}},
  \bibinfo {author} {\bibfnamefont {M.~M.}\ \bibnamefont {Fermann}}, \ and\
  \bibinfo {author} {\bibfnamefont {P.~G.~P.}\ \bibnamefont {Schunemann}},\
  }\href@noop {} {\bibfield  {journal} {\bibinfo  {journal} {Opt. Express}\
  }\textbf {\bibinfo {volume} {20}},\ \bibinfo {pages} {7046} (\bibinfo {year}
  {2012})}\BibitemShut {NoStop}%
\bibitem [{\citenamefont {Keilmann}\ and\ \citenamefont
  {Amarie}(2012)}]{Keilmann:2012bb}%
  \BibitemOpen
  \bibfield  {author} {\bibinfo {author} {\bibfnamefont {F.}~\bibnamefont
  {Keilmann}}\ and\ \bibinfo {author} {\bibfnamefont {S.}~\bibnamefont
  {Amarie}},\ }\href@noop {} {\bibfield  {journal} {\bibinfo  {journal} {J.
  Infrared Millim. Terahertz Waves}\ }\textbf {\bibinfo {volume} {33}},\
  \bibinfo {pages} {479} (\bibinfo {year} {2012})}\BibitemShut {NoStop}%
\bibitem [{\citenamefont {Galli}\ \emph {et~al.}(2013)\citenamefont {Galli},
  \citenamefont {Cappelli}, \citenamefont {Cancio}, \citenamefont {Giusfredi},
  \citenamefont {Mazzotti}, \citenamefont {Bartalini},\ and\ \citenamefont
  {De~Natale}}]{Galli:2013cg}%
  \BibitemOpen
  \bibfield  {author} {\bibinfo {author} {\bibfnamefont {I.}~\bibnamefont
  {Galli}}, \bibinfo {author} {\bibfnamefont {F.}~\bibnamefont {Cappelli}},
  \bibinfo {author} {\bibfnamefont {P.}~\bibnamefont {Cancio}}, \bibinfo
  {author} {\bibfnamefont {G.}~\bibnamefont {Giusfredi}}, \bibinfo {author}
  {\bibfnamefont {D.}~\bibnamefont {Mazzotti}}, \bibinfo {author}
  {\bibfnamefont {S.}~\bibnamefont {Bartalini}}, \ and\ \bibinfo {author}
  {\bibfnamefont {P.}~\bibnamefont {De~Natale}},\ }\href@noop {} {\bibfield
  {journal} {\bibinfo  {journal} {Opt. Express}\ }\textbf {\bibinfo {volume}
  {21}},\ \bibinfo {pages} {28877} (\bibinfo {year} {2013})}\BibitemShut
  {NoStop}%
\bibitem [{\citenamefont {Gambetta}\ \emph {et~al.}(2013)\citenamefont
  {Gambetta}, \citenamefont {Coluccelli}, \citenamefont {Cassinerio},
  \citenamefont {Gatti}, \citenamefont {Laporta}, \citenamefont {Galzerano},\
  and\ \citenamefont {Marangoni}}]{Gambetta:2013ci}%
  \BibitemOpen
  \bibfield  {author} {\bibinfo {author} {\bibfnamefont {A.}~\bibnamefont
  {Gambetta}}, \bibinfo {author} {\bibfnamefont {N.}~\bibnamefont
  {Coluccelli}}, \bibinfo {author} {\bibfnamefont {M.}~\bibnamefont
  {Cassinerio}}, \bibinfo {author} {\bibfnamefont {D.}~\bibnamefont {Gatti}},
  \bibinfo {author} {\bibfnamefont {P.}~\bibnamefont {Laporta}}, \bibinfo
  {author} {\bibfnamefont {G.}~\bibnamefont {Galzerano}}, \ and\ \bibinfo
  {author} {\bibfnamefont {M.}~\bibnamefont {Marangoni}},\ }\href@noop {}
  {\bibfield  {journal} {\bibinfo  {journal} {Opt. Lett.}\ }\textbf {\bibinfo
  {volume} {38}},\ \bibinfo {pages} {1155} (\bibinfo {year}
  {2013})}\BibitemShut {NoStop}%
\bibitem [{\citenamefont {Ru}\ \emph {et~al.}(2017)\citenamefont {Ru},
  \citenamefont {Loparo}, \citenamefont {Zhang}, \citenamefont {Crystal},
  \citenamefont {Vasu}, \citenamefont {Schunemann},\ and\ \citenamefont
  {Vodopyanov}}]{Ru:2017db}%
  \BibitemOpen
  \bibfield  {author} {\bibinfo {author} {\bibfnamefont {Q.}~\bibnamefont
  {Ru}}, \bibinfo {author} {\bibfnamefont {Z.~E.}\ \bibnamefont {Loparo}},
  \bibinfo {author} {\bibfnamefont {X.}~\bibnamefont {Zhang}}, \bibinfo
  {author} {\bibfnamefont {S.}~\bibnamefont {Crystal}}, \bibinfo {author}
  {\bibfnamefont {S.}~\bibnamefont {Vasu}}, \bibinfo {author} {\bibfnamefont
  {P.~G.}\ \bibnamefont {Schunemann}}, \ and\ \bibinfo {author} {\bibfnamefont
  {K.~L.}\ \bibnamefont {Vodopyanov}},\ }\href@noop {} {\bibfield  {journal}
  {\bibinfo  {journal} {Opt. Lett.}\ }\textbf {\bibinfo {volume} {42}},\
  \bibinfo {pages} {4756} (\bibinfo {year} {2017})}\BibitemShut {NoStop}%
\bibitem [{\citenamefont {Kreuzer}(1969)}]{Kreuzer:1969vm}%
  \BibitemOpen
  \bibfield  {author} {\bibinfo {author} {\bibfnamefont {L.~B.}\ \bibnamefont
  {Kreuzer}},\ }in\ \href@noop {} {\emph {\bibinfo {booktitle} {Proceedings of
  the Joint Conference on Lasers and Opto-Electronics}}}\ (\bibinfo
  {publisher} {I.E.R.E.},\ \bibinfo {address} {London},\ \bibinfo {year}
  {1969})\ pp.\ \bibinfo {pages} {52--65}\BibitemShut {NoStop}%
\bibitem [{\citenamefont {Nabors}\ \emph {et~al.}(1990)\citenamefont {Nabors},
  \citenamefont {Yang}, \citenamefont {Day},\ and\ \citenamefont
  {Byer}}]{Nabors:1990vl}%
  \BibitemOpen
  \bibfield  {author} {\bibinfo {author} {\bibfnamefont {C.~D.}\ \bibnamefont
  {Nabors}}, \bibinfo {author} {\bibfnamefont {S.~T.}\ \bibnamefont {Yang}},
  \bibinfo {author} {\bibfnamefont {T.}~\bibnamefont {Day}}, \ and\ \bibinfo
  {author} {\bibfnamefont {R.~L.}\ \bibnamefont {Byer}},\ }\href@noop {}
  {\bibfield  {journal} {\bibinfo  {journal} {J. Opt. Soc. Am. B}\ }\textbf
  {\bibinfo {volume} {7}},\ \bibinfo {pages} {815} (\bibinfo {year}
  {1990})}\BibitemShut {NoStop}%
\bibitem [{\citenamefont {Drummond}\ \emph {et~al.}(1980)\citenamefont
  {Drummond}, \citenamefont {McNeil},\ and\ \citenamefont
  {Walls}}]{Drummond:1980bj}%
  \BibitemOpen
  \bibfield  {author} {\bibinfo {author} {\bibfnamefont {P.~D.}\ \bibnamefont
  {Drummond}}, \bibinfo {author} {\bibfnamefont {K.~J.}\ \bibnamefont
  {McNeil}}, \ and\ \bibinfo {author} {\bibfnamefont {D.~F.}\ \bibnamefont
  {Walls}},\ }\href@noop {} {\bibfield  {journal} {\bibinfo  {journal} {Optica
  Acta}\ }\textbf {\bibinfo {volume} {27}},\ \bibinfo {pages} {321} (\bibinfo
  {year} {1980})}\BibitemShut {NoStop}%
\bibitem [{\citenamefont {Lugiato}\ \emph {et~al.}(1988)\citenamefont
  {Lugiato}, \citenamefont {Oldano}, \citenamefont {Fabre}, \citenamefont
  {Giacobino},\ and\ \citenamefont {Horowicz}}]{Lugiato:1988ep}%
  \BibitemOpen
  \bibfield  {author} {\bibinfo {author} {\bibfnamefont {L.~A.}\ \bibnamefont
  {Lugiato}}, \bibinfo {author} {\bibfnamefont {C.}~\bibnamefont {Oldano}},
  \bibinfo {author} {\bibfnamefont {C.}~\bibnamefont {Fabre}}, \bibinfo
  {author} {\bibfnamefont {E.}~\bibnamefont {Giacobino}}, \ and\ \bibinfo
  {author} {\bibfnamefont {R.~J.}\ \bibnamefont {Horowicz}},\ }\href@noop {}
  {\bibfield  {journal} {\bibinfo  {journal} {Il Nuovo Cimento D}\ }\textbf
  {\bibinfo {volume} {10}},\ \bibinfo {pages} {959} (\bibinfo {year}
  {1988})}\BibitemShut {NoStop}%
\bibitem [{\citenamefont {Oppo}\ \emph
  {et~al.}(1994{\natexlab{a}})\citenamefont {Oppo}, \citenamefont {Brambilla},\
  and\ \citenamefont {Lugiato}}]{Oppo:1994en}%
  \BibitemOpen
  \bibfield  {author} {\bibinfo {author} {\bibfnamefont {G.-L.}\ \bibnamefont
  {Oppo}}, \bibinfo {author} {\bibfnamefont {M.}~\bibnamefont {Brambilla}}, \
  and\ \bibinfo {author} {\bibfnamefont {L.~A.}\ \bibnamefont {Lugiato}},\
  }\href@noop {} {\bibfield  {journal} {\bibinfo  {journal} {Phys. Rev. A}\
  }\textbf {\bibinfo {volume} {49}},\ \bibinfo {pages} {2028} (\bibinfo {year}
  {1994}{\natexlab{a}})}\BibitemShut {NoStop}%
\bibitem [{\citenamefont {Oppo}\ \emph
  {et~al.}(1994{\natexlab{b}})\citenamefont {Oppo}, \citenamefont {Brambilla},
  \citenamefont {Camesasca}, \citenamefont {Gatti},\ and\ \citenamefont
  {Lugiato}}]{Oppo:1994kf}%
  \BibitemOpen
  \bibfield  {author} {\bibinfo {author} {\bibfnamefont {G.-L.}\ \bibnamefont
  {Oppo}}, \bibinfo {author} {\bibfnamefont {M.}~\bibnamefont {Brambilla}},
  \bibinfo {author} {\bibfnamefont {D.}~\bibnamefont {Camesasca}}, \bibinfo
  {author} {\bibfnamefont {A.}~\bibnamefont {Gatti}}, \ and\ \bibinfo {author}
  {\bibfnamefont {L.~A.}\ \bibnamefont {Lugiato}},\ }\href@noop {} {\bibfield
  {journal} {\bibinfo  {journal} {J. Mod. Opt.}\ }\textbf {\bibinfo {volume}
  {41}},\ \bibinfo {pages} {1151} (\bibinfo {year}
  {1994}{\natexlab{b}})}\BibitemShut {NoStop}%
\bibitem [{\citenamefont {de~Valc{\'a}rcel}\ \emph {et~al.}(1996)\citenamefont
  {de~Valc{\'a}rcel}, \citenamefont {Staliunas}, \citenamefont {Rold{\'a}n},\
  and\ \citenamefont {Sanchez-Morcillo}}]{DeValcarcel:1996df}%
  \BibitemOpen
  \bibfield  {author} {\bibinfo {author} {\bibfnamefont {G.~J.}\ \bibnamefont
  {de~Valc{\'a}rcel}}, \bibinfo {author} {\bibfnamefont {K.}~\bibnamefont
  {Staliunas}}, \bibinfo {author} {\bibfnamefont {E.}~\bibnamefont
  {Rold{\'a}n}}, \ and\ \bibinfo {author} {\bibfnamefont {V.~J.}\ \bibnamefont
  {Sanchez-Morcillo}},\ }\href@noop {} {\bibfield  {journal} {\bibinfo
  {journal} {Phys. Rev. A}\ }\textbf {\bibinfo {volume} {54}},\ \bibinfo
  {pages} {1609} (\bibinfo {year} {1996})}\BibitemShut {NoStop}%
\bibitem [{\citenamefont {Staliunas}(1995)}]{Staliunas:1995ky}%
  \BibitemOpen
  \bibfield  {author} {\bibinfo {author} {\bibfnamefont {K.}~\bibnamefont
  {Staliunas}},\ }\href@noop {} {\bibfield  {journal} {\bibinfo  {journal} {J.
  Mod. Opt.}\ }\textbf {\bibinfo {volume} {42}},\ \bibinfo {pages} {1261}
  (\bibinfo {year} {1995})}\BibitemShut {NoStop}%
\bibitem [{\citenamefont {Longhi}(1996{\natexlab{a}})}]{Longhi:1996ft}%
  \BibitemOpen
  \bibfield  {author} {\bibinfo {author} {\bibfnamefont {S.}~\bibnamefont
  {Longhi}},\ }\href@noop {} {\bibfield  {journal} {\bibinfo  {journal} {Phys.
  Rev. A}\ }\textbf {\bibinfo {volume} {53}},\ \bibinfo {pages} {4488}
  (\bibinfo {year} {1996}{\natexlab{a}})}\BibitemShut {NoStop}%
\bibitem [{\citenamefont {Longhi}(1997)}]{Longhi:1997iy}%
  \BibitemOpen
  \bibfield  {author} {\bibinfo {author} {\bibfnamefont {S.}~\bibnamefont
  {Longhi}},\ }\href@noop {} {\bibfield  {journal} {\bibinfo  {journal} {Phys.
  Scripta}\ }\textbf {\bibinfo {volume} {56}},\ \bibinfo {pages} {611}
  (\bibinfo {year} {1997})}\BibitemShut {NoStop}%
\bibitem [{\citenamefont {Longhi}(1995)}]{Longhi:1995jm}%
  \BibitemOpen
  \bibfield  {author} {\bibinfo {author} {\bibfnamefont {S.}~\bibnamefont
  {Longhi}},\ }\href@noop {} {\bibfield  {journal} {\bibinfo  {journal} {Opt.
  Lett.}\ }\textbf {\bibinfo {volume} {20}},\ \bibinfo {pages} {695} (\bibinfo
  {year} {1995})}\BibitemShut {NoStop}%
\bibitem [{\citenamefont {Longhi}(1996{\natexlab{b}})}]{Longhi:1996dh}%
  \BibitemOpen
  \bibfield  {author} {\bibinfo {author} {\bibfnamefont {S.}~\bibnamefont
  {Longhi}},\ }\href@noop {} {\bibfield  {journal} {\bibinfo  {journal} {Opt.
  Lett.}\ }\textbf {\bibinfo {volume} {21}},\ \bibinfo {pages} {860} (\bibinfo
  {year} {1996}{\natexlab{b}})}\BibitemShut {NoStop}%
\bibitem [{\citenamefont {Longhi}(1996{\natexlab{c}})}]{Longhi:1996dm}%
  \BibitemOpen
  \bibfield  {author} {\bibinfo {author} {\bibfnamefont {S.}~\bibnamefont
  {Longhi}},\ }\href@noop {} {\bibfield  {journal} {\bibinfo  {journal} {J.
  Mod. Opt.}\ }\textbf {\bibinfo {volume} {43}},\ \bibinfo {pages} {1089}
  (\bibinfo {year} {1996}{\natexlab{c}})}\BibitemShut {NoStop}%
\bibitem [{\citenamefont {Longhi}(1996{\natexlab{d}})}]{Longhi:1996gt}%
  \BibitemOpen
  \bibfield  {author} {\bibinfo {author} {\bibfnamefont {S.}~\bibnamefont
  {Longhi}},\ }\href@noop {} {\bibfield  {journal} {\bibinfo  {journal} {Phys.
  Rev. E}\ }\textbf {\bibinfo {volume} {53}},\ \bibinfo {pages} {5520}
  (\bibinfo {year} {1996}{\natexlab{d}})}\BibitemShut {NoStop}%
\bibitem [{\citenamefont {Boyd}(2008)}]{Boyd:NLO}%
  \BibitemOpen
  \bibfield  {author} {\bibinfo {author} {\bibfnamefont {R.~W.}\ \bibnamefont
  {Boyd}},\ }\href@noop {} {\emph {\bibinfo {title} {Nonlinear Optics}}},\
  \bibinfo {edition} {3rd}\ ed.\ (\bibinfo  {publisher} {Academic Press},\
  \bibinfo {address} {Burlington},\ \bibinfo {year} {2008})\BibitemShut
  {NoStop}%
\bibitem [{\citenamefont {Leo}\ \emph {et~al.}(2016{\natexlab{a}})\citenamefont
  {Leo}, \citenamefont {Hansson}, \citenamefont {Ricciardi}, \citenamefont
  {De~Rosa}, \citenamefont {Coen}, \citenamefont {Wabnitz},\ and\ \citenamefont
  {Erkintalo}}]{Leo:2016kj}%
  \BibitemOpen
  \bibfield  {author} {\bibinfo {author} {\bibfnamefont {F.}~\bibnamefont
  {Leo}}, \bibinfo {author} {\bibfnamefont {T.}~\bibnamefont {Hansson}},
  \bibinfo {author} {\bibfnamefont {I.}~\bibnamefont {Ricciardi}}, \bibinfo
  {author} {\bibfnamefont {M.}~\bibnamefont {De~Rosa}}, \bibinfo {author}
  {\bibfnamefont {S.}~\bibnamefont {Coen}}, \bibinfo {author} {\bibfnamefont
  {S.}~\bibnamefont {Wabnitz}}, \ and\ \bibinfo {author} {\bibfnamefont
  {M.}~\bibnamefont {Erkintalo}},\ }\href@noop {} {\bibfield  {journal}
  {\bibinfo  {journal} {Phys. Rev. Lett.}\ }\textbf {\bibinfo {volume} {116}},\
  \bibinfo {pages} {{033901}} (\bibinfo {year}
  {2016}{\natexlab{a}})}\BibitemShut {NoStop}%
\bibitem [{Note1()}]{Note1}%
  \BibitemOpen
  \bibinfo {note} {Supplemental Material at ... includes the detailed
  derivation and the stability analysis.}\BibitemShut {Stop}%
\bibitem [{\citenamefont {Kibler}\ \emph {et~al.}(2012)\citenamefont {Kibler},
  \citenamefont {Michel}, \citenamefont {Garnier},\ and\ \citenamefont
  {Picozzi}}]{Kibler:2012ho}%
  \BibitemOpen
  \bibfield  {author} {\bibinfo {author} {\bibfnamefont {B.}~\bibnamefont
  {Kibler}}, \bibinfo {author} {\bibfnamefont {C.}~\bibnamefont {Michel}},
  \bibinfo {author} {\bibfnamefont {J.}~\bibnamefont {Garnier}}, \ and\
  \bibinfo {author} {\bibfnamefont {A.}~\bibnamefont {Picozzi}},\ }\href@noop
  {} {\bibfield  {journal} {\bibinfo  {journal} {Opt. Lett.}\ }\textbf
  {\bibinfo {volume} {37}},\ \bibinfo {pages} {2472} (\bibinfo {year}
  {2012})}\BibitemShut {NoStop}%
\bibitem [{\citenamefont {Hansson}\ \emph {et~al.}(2014)\citenamefont
  {Hansson}, \citenamefont {Modotto},\ and\ \citenamefont
  {Wabnitz}}]{Hansson:2014do}%
  \BibitemOpen
  \bibfield  {author} {\bibinfo {author} {\bibfnamefont {T.}~\bibnamefont
  {Hansson}}, \bibinfo {author} {\bibfnamefont {D.}~\bibnamefont {Modotto}}, \
  and\ \bibinfo {author} {\bibfnamefont {S.}~\bibnamefont {Wabnitz}},\
  }\href@noop {} {\bibfield  {journal} {\bibinfo  {journal} {Opt. Lett.}\
  }\textbf {\bibinfo {volume} {39}},\ \bibinfo {pages} {6747} (\bibinfo {year}
  {2014})}\BibitemShut {NoStop}%
\bibitem [{\citenamefont {Okawachi}\ \emph {et~al.}(2017)\citenamefont
  {Okawachi}, \citenamefont {Yu}, \citenamefont {Venkataraman}, \citenamefont
  {Latawiec}, \citenamefont {Griffith}, \citenamefont {Lipson}, \citenamefont
  {Lon{\v c}ar},\ and\ \citenamefont {Gaeta}}]{Okawachi:2017jl}%
  \BibitemOpen
  \bibfield  {author} {\bibinfo {author} {\bibfnamefont {Y.}~\bibnamefont
  {Okawachi}}, \bibinfo {author} {\bibfnamefont {M.}~\bibnamefont {Yu}},
  \bibinfo {author} {\bibfnamefont {V.}~\bibnamefont {Venkataraman}}, \bibinfo
  {author} {\bibfnamefont {P.~M.}\ \bibnamefont {Latawiec}}, \bibinfo {author}
  {\bibfnamefont {A.~G.}\ \bibnamefont {Griffith}}, \bibinfo {author}
  {\bibfnamefont {M.}~\bibnamefont {Lipson}}, \bibinfo {author} {\bibfnamefont
  {M.}~\bibnamefont {Lon{\v c}ar}}, \ and\ \bibinfo {author} {\bibfnamefont
  {A.~L.}\ \bibnamefont {Gaeta}},\ }\href@noop {} {\bibfield  {journal}
  {\bibinfo  {journal} {Opt. Lett.}\ }\textbf {\bibinfo {volume} {42}},\
  \bibinfo {pages} {2786} (\bibinfo {year} {2017})}\BibitemShut {NoStop}%
\bibitem [{\citenamefont {Hansson}\ \emph {et~al.}(2017)\citenamefont
  {Hansson}, \citenamefont {Leo}, \citenamefont {Erkintalo}, \citenamefont
  {Coen}, \citenamefont {Ricciardi}, \citenamefont {De~Rosa},\ and\
  \citenamefont {Wabnitz}}]{Hansson:2017cs}%
  \BibitemOpen
  \bibfield  {author} {\bibinfo {author} {\bibfnamefont {T.}~\bibnamefont
  {Hansson}}, \bibinfo {author} {\bibfnamefont {F.}~\bibnamefont {Leo}},
  \bibinfo {author} {\bibfnamefont {M.}~\bibnamefont {Erkintalo}}, \bibinfo
  {author} {\bibfnamefont {S.}~\bibnamefont {Coen}}, \bibinfo {author}
  {\bibfnamefont {I.}~\bibnamefont {Ricciardi}}, \bibinfo {author}
  {\bibfnamefont {M.}~\bibnamefont {De~Rosa}}, \ and\ \bibinfo {author}
  {\bibfnamefont {S.}~\bibnamefont {Wabnitz}},\ }\href@noop {} {\bibfield
  {journal} {\bibinfo  {journal} {Phys. Rev. A}\ }\textbf {\bibinfo {volume}
  {95}},\ \bibinfo {pages} {013805} (\bibinfo {year} {2017})}\BibitemShut
  {NoStop}%
\bibitem [{\citenamefont {Staliunas}\ and\ \citenamefont
  {Sanchez-Morcillo}(2000)}]{Staliunas:2000tu}%
  \BibitemOpen
  \bibfield  {author} {\bibinfo {author} {\bibfnamefont {K.}~\bibnamefont
  {Staliunas}}\ and\ \bibinfo {author} {\bibfnamefont {V.~J.}\ \bibnamefont
  {Sanchez-Morcillo}},\ }\href@noop {} {\bibfield  {journal} {\bibinfo
  {journal} {Opt. Commun.}\ }\textbf {\bibinfo {volume} {177}},\ \bibinfo
  {pages} {389} (\bibinfo {year} {2000})}\BibitemShut {NoStop}%
\bibitem [{\citenamefont {Bortolozzo}\ \emph {et~al.}(2001)\citenamefont
  {Bortolozzo}, \citenamefont {Villoresi},\ and\ \citenamefont
  {Ramazza}}]{Bortolozzo:2001dt}%
  \BibitemOpen
  \bibfield  {author} {\bibinfo {author} {\bibfnamefont {U.}~\bibnamefont
  {Bortolozzo}}, \bibinfo {author} {\bibfnamefont {P.}~\bibnamefont
  {Villoresi}}, \ and\ \bibinfo {author} {\bibfnamefont {P.~L.}\ \bibnamefont
  {Ramazza}},\ }\href@noop {} {\bibfield  {journal} {\bibinfo  {journal} {Phys.
  Rev. Lett.}\ }\textbf {\bibinfo {volume} {87}},\ \bibinfo {pages} {274102}
  (\bibinfo {year} {2001})}\BibitemShut {NoStop}%
\bibitem [{\citenamefont {Esteban-Mart{\'\i}n}\ \emph
  {et~al.}(2004)\citenamefont {Esteban-Mart{\'\i}n}, \citenamefont
  {Garc{\'\i}a}, \citenamefont {Rold{\'a}n}, \citenamefont {Taranenko},
  \citenamefont {{de Valc{\'a}rcel}},\ and\ \citenamefont
  {Weiss}}]{EstebanMartin:2004dv}%
  \BibitemOpen
  \bibfield  {author} {\bibinfo {author} {\bibfnamefont {A.}~\bibnamefont
  {Esteban-Mart{\'\i}n}}, \bibinfo {author} {\bibfnamefont {J.}~\bibnamefont
  {Garc{\'\i}a}}, \bibinfo {author} {\bibfnamefont {E.}~\bibnamefont
  {Rold{\'a}n}}, \bibinfo {author} {\bibfnamefont {V.~B.}\ \bibnamefont
  {Taranenko}}, \bibinfo {author} {\bibfnamefont {G.~J.}\ \bibnamefont {{de
  Valc{\'a}rcel}}}, \ and\ \bibinfo {author} {\bibfnamefont {C.~O.}\
  \bibnamefont {Weiss}},\ }\href@noop {} {\bibfield  {journal} {\bibinfo
  {journal} {Phys. Rev. A}\ }\textbf {\bibinfo {volume} {69}},\ \bibinfo
  {pages} {033816} (\bibinfo {year} {2004})}\BibitemShut {NoStop}%
\bibitem [{\citenamefont {Drever}\ \emph {et~al.}(1983)\citenamefont {Drever},
  \citenamefont {Hall}, \citenamefont {Kowalski}, \citenamefont {Hough},
  \citenamefont {Ford}, \citenamefont {Munley},\ and\ \citenamefont
  {Ward}}]{Drever:1983gx}%
  \BibitemOpen
  \bibfield  {author} {\bibinfo {author} {\bibfnamefont {R.~W.~P.}\
  \bibnamefont {Drever}}, \bibinfo {author} {\bibfnamefont {J.~L.}\
  \bibnamefont {Hall}}, \bibinfo {author} {\bibfnamefont {F.~V.}\ \bibnamefont
  {Kowalski}}, \bibinfo {author} {\bibfnamefont {J.}~\bibnamefont {Hough}},
  \bibinfo {author} {\bibfnamefont {G.~M.}\ \bibnamefont {Ford}}, \bibinfo
  {author} {\bibfnamefont {A.~J.}\ \bibnamefont {Munley}}, \ and\ \bibinfo
  {author} {\bibfnamefont {H.}~\bibnamefont {Ward}},\ }\href@noop {} {\bibfield
   {journal} {\bibinfo  {journal} {Appl. Phys. B}\ }\textbf {\bibinfo {volume}
  {31}},\ \bibinfo {pages} {97} (\bibinfo {year} {1983})}\BibitemShut {NoStop}%
\bibitem [{\citenamefont {Ricciardi}\ \emph {et~al.}(2010)\citenamefont
  {Ricciardi}, \citenamefont {De~Rosa}, \citenamefont {Rocco}, \citenamefont
  {Ferraro},\ and\ \citenamefont {De~Natale}}]{Ricciardi:2010kd}%
  \BibitemOpen
  \bibfield  {author} {\bibinfo {author} {\bibfnamefont {I.}~\bibnamefont
  {Ricciardi}}, \bibinfo {author} {\bibfnamefont {M.}~\bibnamefont {De~Rosa}},
  \bibinfo {author} {\bibfnamefont {A.}~\bibnamefont {Rocco}}, \bibinfo
  {author} {\bibfnamefont {P.}~\bibnamefont {Ferraro}}, \ and\ \bibinfo
  {author} {\bibfnamefont {P.}~\bibnamefont {De~Natale}},\ }\href@noop {}
  {\bibfield  {journal} {\bibinfo  {journal} {Opt. Express}\ }\textbf {\bibinfo
  {volume} {18}},\ \bibinfo {pages} {10985} (\bibinfo {year}
  {2010})}\BibitemShut {NoStop}%
\bibitem [{\citenamefont {Leo}\ \emph {et~al.}(2016{\natexlab{b}})\citenamefont
  {Leo}, \citenamefont {Hansson}, \citenamefont {Ricciardi}, \citenamefont
  {De~Rosa}, \citenamefont {Coen}, \citenamefont {Wabnitz},\ and\ \citenamefont
  {Erkintalo}}]{Leo:2016df}%
  \BibitemOpen
  \bibfield  {author} {\bibinfo {author} {\bibfnamefont {F.}~\bibnamefont
  {Leo}}, \bibinfo {author} {\bibfnamefont {T.}~\bibnamefont {Hansson}},
  \bibinfo {author} {\bibfnamefont {I.}~\bibnamefont {Ricciardi}}, \bibinfo
  {author} {\bibfnamefont {M.}~\bibnamefont {De~Rosa}}, \bibinfo {author}
  {\bibfnamefont {S.}~\bibnamefont {Coen}}, \bibinfo {author} {\bibfnamefont
  {S.}~\bibnamefont {Wabnitz}}, \ and\ \bibinfo {author} {\bibfnamefont
  {M.}~\bibnamefont {Erkintalo}},\ }\href@noop {} {\bibfield  {journal}
  {\bibinfo  {journal} {Phys. Rev. A}\ }\textbf {\bibinfo {volume} {93}},\
  \bibinfo {pages} {043831} (\bibinfo {year} {2016}{\natexlab{b}})}\BibitemShut
  {NoStop}%
\bibitem [{\citenamefont {Parisi}\ \emph {et~al.}(2017)\citenamefont {Parisi},
  \citenamefont {Morais}, \citenamefont {Ricciardi}, \citenamefont {Mosca},
  \citenamefont {Hansson}, \citenamefont {Wabnitz}, \citenamefont {Leo},\ and\
  \citenamefont {De~Rosa}}]{Parisi:2017fp}%
  \BibitemOpen
  \bibfield  {author} {\bibinfo {author} {\bibfnamefont {M.}~\bibnamefont
  {Parisi}}, \bibinfo {author} {\bibfnamefont {N.}~\bibnamefont {Morais}},
  \bibinfo {author} {\bibfnamefont {I.}~\bibnamefont {Ricciardi}}, \bibinfo
  {author} {\bibfnamefont {S.}~\bibnamefont {Mosca}}, \bibinfo {author}
  {\bibfnamefont {T.}~\bibnamefont {Hansson}}, \bibinfo {author} {\bibfnamefont
  {S.}~\bibnamefont {Wabnitz}}, \bibinfo {author} {\bibfnamefont
  {G.}~\bibnamefont {Leo}}, \ and\ \bibinfo {author} {\bibfnamefont
  {M.}~\bibnamefont {De~Rosa}},\ }\href@noop {} {\bibfield  {journal} {\bibinfo
   {journal} {J. Opt. Soc. Am. B}\ }\textbf {\bibinfo {volume} {34}},\ \bibinfo
  {pages} {1842} (\bibinfo {year} {2017})}\BibitemShut {NoStop}%
\bibitem [{\citenamefont {Erkintalo}\ and\ \citenamefont
  {Coen}(2014)}]{Erkintalo:2014fq}%
  \BibitemOpen
  \bibfield  {author} {\bibinfo {author} {\bibfnamefont {M.}~\bibnamefont
  {Erkintalo}}\ and\ \bibinfo {author} {\bibfnamefont {S.}~\bibnamefont
  {Coen}},\ }\href@noop {} {\bibfield  {journal} {\bibinfo  {journal} {Opt.
  Lett.}\ }\textbf {\bibinfo {volume} {39}},\ \bibinfo {pages} {283} (\bibinfo
  {year} {2014})}\BibitemShut {NoStop}%
\bibitem [{\citenamefont {Wen}\ \emph {et~al.}(2016)\citenamefont {Wen},
  \citenamefont {Lamont}, \citenamefont {Strogatz},\ and\ \citenamefont
  {Gaeta}}]{Wen:2016cz}%
  \BibitemOpen
  \bibfield  {author} {\bibinfo {author} {\bibfnamefont {Y.~H.}\ \bibnamefont
  {Wen}}, \bibinfo {author} {\bibfnamefont {M.~R.~E.}\ \bibnamefont {Lamont}},
  \bibinfo {author} {\bibfnamefont {S.~H.}\ \bibnamefont {Strogatz}}, \ and\
  \bibinfo {author} {\bibfnamefont {A.~L.}\ \bibnamefont {Gaeta}},\ }\href@noop
  {} {\bibfield  {journal} {\bibinfo  {journal} {Phys. Rev. A}\ }\textbf
  {\bibinfo {volume} {94}},\ \bibinfo {pages} {063843} (\bibinfo {year}
  {2016})}\BibitemShut {NoStop}%
\bibitem [{\citenamefont {Morais}\ \emph {et~al.}(2017)\citenamefont {Morais},
  \citenamefont {Roland}, \citenamefont {Ravaro}, \citenamefont {Hease},
  \citenamefont {Lema{\^\i}tre}, \citenamefont {Gomez}, \citenamefont
  {Wabnitz}, \citenamefont {De~Rosa}, \citenamefont {Favero},\ and\
  \citenamefont {Leo}}]{Morais:2017va}%
  \BibitemOpen
  \bibfield  {author} {\bibinfo {author} {\bibfnamefont {N.}~\bibnamefont
  {Morais}}, \bibinfo {author} {\bibfnamefont {I.}~\bibnamefont {Roland}},
  \bibinfo {author} {\bibfnamefont {M.}~\bibnamefont {Ravaro}}, \bibinfo
  {author} {\bibfnamefont {W.}~\bibnamefont {Hease}}, \bibinfo {author}
  {\bibfnamefont {A.}~\bibnamefont {Lema{\^\i}tre}}, \bibinfo {author}
  {\bibfnamefont {C.}~\bibnamefont {Gomez}}, \bibinfo {author} {\bibfnamefont
  {S.}~\bibnamefont {Wabnitz}}, \bibinfo {author} {\bibfnamefont
  {M.}~\bibnamefont {De~Rosa}}, \bibinfo {author} {\bibfnamefont
  {I.}~\bibnamefont {Favero}}, \ and\ \bibinfo {author} {\bibfnamefont
  {G.}~\bibnamefont {Leo}},\ }\href@noop {} {\bibfield  {journal} {\bibinfo
  {journal} {Opt. Lett.}\ }\textbf {\bibinfo {volume} {42}},\ \bibinfo {pages}
  {4287} (\bibinfo {year} {2017})}\BibitemShut {NoStop}%
\bibitem [{\citenamefont {Timurdogan}\ \emph {et~al.}(2017)\citenamefont
  {Timurdogan}, \citenamefont {Poulton}, \citenamefont {Byrd},\ and\
  \citenamefont {Watts}}]{Timurdogan:2017jg}%
  \BibitemOpen
  \bibfield  {author} {\bibinfo {author} {\bibfnamefont {E.}~\bibnamefont
  {Timurdogan}}, \bibinfo {author} {\bibfnamefont {C.~V.}\ \bibnamefont
  {Poulton}}, \bibinfo {author} {\bibfnamefont {M.~J.}\ \bibnamefont {Byrd}}, \
  and\ \bibinfo {author} {\bibfnamefont {M.~R.}\ \bibnamefont {Watts}},\
  }\href@noop {} {\bibfield  {journal} {\bibinfo  {journal} {Nature Photon.}\
  }\textbf {\bibinfo {volume} {11}},\ \bibinfo {pages} {200} (\bibinfo {year}
  {2017})}\BibitemShut {NoStop}%
\end{thebibliography}

\begin{thebibliography}{2}

\bibitem{Leo:2016kj}
F. Leo, T. Hansson, I. Ricciardi, M. De Rosa, S. Coen, S.~Wabnitz, and M.~Erkintalo,
Phys. Rev. Lett.  \textbf{116}, 033901 (2016).

\bibitem{Leo:2016df}
F. Leo, T. Hansson, I. Ricciardi, M. De Rosa, S. Coen, S.~Wabnitz, and M.~Erkintalo, 
Phys. Rev. A \textbf{93}, 043831 (2016).

\bibitem{Hansson:2017cs}
T.~Hansson, F.~Leo, M.~Erkintalo, S.~Coen, I.~Ricciardi, M.~De~Rosa,  and S.~Wabnitz,
Phys. Rev. A \textbf{95}, 013805 (2017).

\end{thebibliography}
\end{document}